\documentclass[reprint,amsmath,notitlepage,amssymb,aps,nofootinbib,superscriptaddress,onecolumn]{revtex4-1}
\usepackage{graphicx}
\usepackage{dcolumn}
\usepackage{bm}
\usepackage[normalem]{ulem}
\usepackage{color}
\usepackage{xcolor}
\usepackage{varioref}
\usepackage{mathrsfs}
\usepackage{amsfonts}
\usepackage{varioref}
\usepackage{csquotes}
\usepackage{hyphenat}
\usepackage{float}
\usepackage{multirow}

\RequirePackage[colorlinks,citecolor=blue,urlcolor=magenta,linkcolor=blue]{hyperref}
\allowdisplaybreaks

\labelformat{section}{Section #1} 
\labelformat{subsection}{Section #1} 
\labelformat{subsubsection}{Section #1}
\labelformat{subsubsubsection}{Section #1}
\labelformat{equation}{Eq.~(#1)} 
\labelformat{figure}{Fig.~#1} 
\labelformat{subfigure}{Fig.~\thefigure#1} 
\labelformat{table}{Table~#1} 
\labelformat{appendix}{Appendix #1}

\begin{document}

\title{Transition from inspiral to plunge for braneworld EMRI}

\author{Sajal Mukherjee}
\email{Present address: Department of Physics, Birla Institute of Technology and Science - Pilani, Rajasthan 333031, India \\
email: sajal.mukherjee@pilani.bits-pilani.ac.in}
\affiliation{Astronomical Institue of the Czech Academy of Sciences, Bocni II 1401/1a, CZ-14100 Prague, Czech Republic}

\author{Sumanta Chakraborty}
\email{tpsc@iacs.res.in}
\affiliation{School of Physical Sciences, Indian Association for the Cultivation of Science, Kolkata-700032, India}

\date{\today}

\begin{abstract}

In the present article, we discuss the late inspiral and then the transition regime to the plunge phase of a secondary, less massive compact object into a more massive braneworld black hole, in the context of an extreme-mass-ratio inspiral. We obtain the approximate expressions for fluxes due to slowly evolving constants of motion, such as the energy and the angular momentum, in the presence of the tidal charge inherited from the higher spacetime dimensions for an extreme-mass-ratio system. These expressions for fluxes are further used to introduce dissipative effects while modeling the inspiral to the plunge phase through the transition regime. Within our setup, we provide a qualitative understanding of how the additional tidal charge present in the braneworld scenario may affect the timescale of the late inspiral to the plunge, in particular, by enhancing the time scale of the transition regime. Finally, we provide an estimate for the tidal charge from the higher dimensions, using the observable aspects of the transition regime from the late inspiral to the plunge by the gravitational wave detectors. 

\end{abstract}
\maketitle

\section{Introduction}\label{BY_Intro}
The overwhelming success story of general relativity (GR) is prospered with the addition of recent observational predictions, especially in terms of the gravitational wave (GW) astronomy \cite{TheLIGOScientific:2016pea,TheLIGOScientific:2017qsa,Abbott:2017gyy} and black hole shadow \cite{EventHorizonTelescope:2019dse,EventHorizonTelescope:2019uob}. The GWs emitted during the merger of two compact objects hold rich information about the background spacetime and hence about the theory of gravity under consideration \cite{Berti:2015itd,Berti:2019xgr,Will:2005va}. So far, all the GW signals originating from such merger events are consistent with the predictions from general relativity \cite{LIGOScientific:2019fpa,LIGOScientific:2020tif}. The same holds true for the shadow measurements of M87* and SgrA* black holes by the Event Horizon Telescope, whose findings are also consistent with general relativity \cite{EventHorizonTelescope:2019jan,EventHorizonTelescope:2019ths}. However, the black hole solutions in general relativity are riddled with singularities \cite{Penrose:1964wq,Hawking:1976ra}, the existence of the Cauchy horizon threatens the deterministic nature of the field equations \cite{Cardoso:2017soq,Rahman:2018oso}, and finally, the information loss paradox \cite{Hawking:1975vcx,Chakraborty:2017pmn,Mathur:2009hf} questions the validity of the equivalence principle. All of these suggestions motivate us to look for alternatives to general relativity, or, maybe alternatives to black holes \cite{Cardoso:2016rao,Wang:2019rcf}. 

Here, we will consider one particular approach, namely the braneworld scenario, which is an alternative to general relativity \cite{Kanti:2004nr,Maartens:2003tw}. In this theory, the compact objects are quantum corrected and hence need not satisfy the properties of a classical black hole spacetime \cite{Emparan:2002px,Emparan:1999wa,Garriga:1999yh}. The modifications over and above Einstein's equations in the vacuum are due to the projected bulk Weyl tensor, i.e., the presence of a non-trivial bulk geometry alters the gravitational field as seen by a four-dimensional observer \cite{Shiromizu:1999wj,Chamblin:2000ra,Dadhich:2000am}. If such modifications can be probed, they will provide tantalizing evidence in favour of the existence of extra spatial dimensions. Moreover, such extra dimensions are often endowed with a negative cosmological constant, and hence by the AdS/CFT correspondence, there will be non-trivial effects due to the quantum CFTs on the brane. All of these modifications lead to corrections to the background geometry and also modify the near horizon behaviour of the spacetime. Implications on the stability of the braneworld geometry, due to such modified near horizon behaviour, for both rotating and non-rotating spacetimes have already been explored \cite{Toshmatov:2016bsb,Dey:2020lhq,Dey:2020pth,Chakraborty:2022zlq,Mishra:2021waw,Andriot:2017oaz}. Further, the effect of such corrections on the tidal properties of compact objects, as well as on the quasi-normal mode frequencies, have been computed and contrasted with the real GW data from GW150914 and GW170817 \cite{Chakraborty:2017qve,Chakravarti:2018vlt,Chakravarti:2019aup}. However, all of these results are in the context of binaries, having equal/nearly-equal mass ratios. In what follows, we will concentrate on the binaries having extreme mass ratios and in particular, we will consider the plunge phase, which joins the inspiral phase to the merger phase. For, equal/nearly-equal mass ratio binaries, this phase is very short-lived and hence is not of much importance, but for binary systems with extreme mass ratio, this phase will have a significant imprint on the GW waveform. The presence of an extra spatial dimension will also affect the inspiral phase and may depict a significant departure from the prediction of general relativity, at least for extreme mass ratio binaries. However, an accurate description of the inspiral phase of any such extreme mass ratio binary also involves various subtle effects, e.g., the self-force. Though these effects are well understood in general relativity, their behaviour in the presence of extra spatial dimension has not been attempted before and is a challenging task to perform. Thus, in this work we concentrate on the plunging phase of an extreme mass ratio binary system, with an extra spatial dimension, rather than studying the inspiral phase. Moreover, the plunging phase happens close to the horizon, and hence one expects the corrections from the higher spatial dimension to be of importance. It is worthwhile to mention that these braneworld models are favoured in the context of black hole shadow measurement, however, the existence of large uncertainties in the measurement of the shadow radius, hinders any stronger predictions \cite{Banerjee:2022jog,Banerjee:2019nnj}.    

Till now, we have observed numerous merger events between compact objects through the ground-based GW detector, namely, the LIGO-Virgo collaboration \cite{ligodoc5}. These events are characterized by equal or nearly equal mass ratios in which the GW frequency falls in the Hz to kHz range. In order to probe the other spectrum, e.g., the merger of compact objects having extreme/intermediate mass ratios, which requires the GW detectors to be sensitive to the frequencies ranging from mHz to Hz \cite{Punturo:2010zza,Evans:2016mbw,Isoyama:2018rjb,Baibhav:2019rsa}. In that regard, the Laser Interferometer Space Antenna (LISA), a future space-based GW detector, is going to detect such low-frequency GW signal, typically radiated in the binary systems undergoing extreme mass ratio inspiral (EMRI) \cite{LISA:2017pwj,Berry:2019wgg}. These GW sources are extremely interesting in their own right, since they have a prolonged inspiral phase there are many subtle effects which can be observed through them, e.g., gravitational self-force \cite{Barack:2009ux}. In addition, due to this long timescale involved with the inspiral, the modelling of gravitational waves emitted in these systems becomes extremely challenging, and some of the recent findings in this field can be found in Refs. \cite{Barack:2018yvs,Isoyama:2021jjd,Warburton:2021kwk}. In the present article, as already emphasized, we are interested in modelling an EMRI at its final stage, i.e., inspiral to plunge \cite{Ori:2000zn,Apte:2019txp,Burke:2019yek}. Due to the proximity of the infalling smaller mass compact object to the horizon of the supermassive object and the significant duration spent in this plunge phase, it is expected that this phase will hold important information about the nature of geometry in the strong gravity regime. Thus we wish to explore, the possibility of distinguishing a spacetime inheriting an extra spatial dimension from the usual four-dimensional spacetime using the plunge phase in an EMRI system.  

The paper is organized as follows: We start \ref{rotbhbrane} with the geometrical aspects of the rotating BH solution on the brane and then discuss the properties of circular geodesics in this spacetime. Subsequently, in \ref{sec:FLUXES}, we compute the fluxes of the energy, the angular momentum, and the Carter constant for an inspiralling secondary compact object, treated as a test particle in the spacetime of the more massive primary object, described by the braneworld BH. Using these fluxes, we compute the trajectory of the secondary object as it progresses from the late inspiral to the plunge through a transition regime, in \ref{transitionbrane} and then also point out the observational implications. We conclude with a discussion of the results obtained. Several derivations have been reserved for the appendices, namely \ref{AppA} to \ref{Appd}. 

\emph{Notations and Conventions:} Throughout this work, we set the fundamental constants, namely $G$ and $c$ to unity. The Greek indices, $\mu,\nu,\cdots$ denote the four-dimensional spacetime coordinates. We follow the mostly positive signature convention, e.g., the four-dimensional flat metric will take the form, $\eta_{\mu \nu}=\textrm{diag.}(-1,+1,+1,+1)$. 
 
\section{Rotating black hole on the brane}\label{rotbhbrane}

In this section, we will briefly review the basic properties of the rotating black hole solution on the brane and then we will explore the nature of circular orbits in this spacetime. Since our analysis will be in a region, which is away from the horizon and will depend only on the geometry of the exterior spacetime, without any reference to the horizon properties, it does not matter if we consider the compact object as a classical or, a quantum black hole. The construction of the rotating metric starts from the effective gravitational field equations on the four-dimensional brane hypersurface, embedded in a five-dimensional spacetime, namely the bulk. The effective field equations are obtained by projecting the bulk Einstein's equations on the lower dimensional brane hypersurface and in this process, non-trivial bulk physics gets imprinted on the brane. Such that, the gravitational dynamics on the vacuum brane gets described by \cite{Shiromizu:1999wj}, 
\begin{align}
~^{(4)}G_{\mu \nu}+E_{\mu \nu}=0~,
\end{align}
where $~^{(4)}G_{\mu \nu}$ is the Einstein tensor constructed using brane geometry alone and $E_{\mu \nu}$ is the projection of the bulk Weyl tensor on the brane hypersurface. Thus the bulk dynamics makes its presence felt on the brane, through this tensor $E_{\mu \nu}$. Given the traceless nature of $E_{\mu \nu}$, owing to the symmetry properties of the Weyl tensor, one can assume this term to be equivalent to the energy-momentum tensor of a Maxwell field, with an overall negative sign \cite{Aliev:2005bi,Harko:2004ui}. The Maxwell charge, e.g., would be related to the length of the extra dimension, through the bulk geometry. In the presence of static and spherical symmetry, the correction to the $g_{tt}$ and $g^{rr}$ components of the Schwarzschild spacetime, due to the projection of the bulk Weyl tensor, is of the form $qM^{2}/r^{2}$, where $q$ is a dimensionless constant, which can be negative as well. While, in the context of axisymmetry, the metric of a rotating black hole on the brane, is presented by the following line element \cite{Aliev:2009cg},
\begin{align}\label{rot_brane}
ds^{2}&=-\left(1-\frac{2Mr-qM^{2}}{\rho^{2}}\right)dt^{2}-\frac{2a\sin^{2}\theta \left(2Mr-qM^{2}\right)}{\rho^{2}}dtd\phi+\rho^{2}\left(\frac{dr^{2}}{\Delta}+d\theta^{2}\right)
\nonumber
\\
&\hskip 4 cm +\left\{r^{2}+a^{2}+\frac{a^{2}\sin^{2}\theta\left(2Mr-qM^{2}\right)}{\rho^{2}} \right\}\sin^{2}\theta d\phi^{2}~.
\end{align}
Here, in analogy with the Kerr spacetime, we have defined, $\rho^{2}\equiv r^{2}+a^{2}\cos^{2}\theta$ and $\Delta\equiv r^{2}-2Mr+a^{2}+qM^{2}$. The main difference of the above line element with that of the Kerr-Newman spacetime is that the charge term $q$ can take negative values. In particular, for positive values of $q$ one recovers the Kerr-Newman spacetime from \ref{rot_brane}, by substituting, $qM^{2}\rightarrow Q^{2}$. Normally, astrophysical black holes cannot have electric charge, either due to charge neutralization with surrounding plasma, or, due to novel shielding effect \cite{Feng:2022evy}, but since the origin of $q$ is from the bulk geometry, the presence of extra dimension can lead to non-zero and negative values for $q$ \cite{Dadhich:2000am}. 

Since we wish to study EMRI, circular orbits play the most important role in this analysis. In what follows, we will first determine the geodesic equations on the equatorial plane and then shall restrict ourselves to the circular orbits. In particular, we are interested in the inner-most stable circular orbit (ISCO), as the smaller mass black hole, after crossing the ISCO, plunges into the higher mass black hole. For this purpose, we note that the rotating spacetime on the brane has two Killing vectors, $(\partial/\partial t)^{\mu}$ and $(\partial/\partial \phi)^{\mu}$, owing to the fact that the metric elements in \ref{rot_brane} neither depend on $t$, nor on $\phi$. As a consequence, the energy and the angular momentum of the geodesics must be conserved. On the equatorial plane, the conserved energy and the conserved angular momentum, per unit mass, can be expressed as, 
\begin{align}
E&=\left(1-\frac{2M}{r}+\frac{qM^{2}}{r^{2}}\right)\dot{t}+a\left(\frac{2M}{r}-\frac{qM^{2}}{r^{2}}\right)\dot{\phi}~,
\label{brane_E}
\\
L_{z}&=-a\left(\frac{2M}{r}-\frac{qM^{2}}{r^{2}}\right)\dot{t}+\left\{r^{2}+a^{2}+a^{2}\left(\frac{2M}{r}-\frac{qM^{2}}{r^{2}}\right)\right\}\dot{\phi}~.
\label{brane_L}
\end{align}
Note that on setting $q=0$, we will get back the respective expressions for energy and angular momentum of geodesics on the equatorial plane of Kerr spacetime. One can also invert \ref{brane_E} and \ref{brane_L}, in order to express the velocities $\dot{t}$ and $\dot{\phi}$ in terms of the metric parameters and the energy $E$ and the angular momentum $L_{z}$. Moreover, the radial motion of the geodesics on the equatorial plane will be governed by, 
\begin{align}
\dot{r}^{2}=\frac{\Delta}{r^{2}}\Bigg[-\delta _{1}+\frac{r^{2}+a^{2}+a^{2}\left(\frac{2M}{r}-\frac{qM^{2}}{r^{2}}\right)}{\Delta}E^{2}
-\frac{2a\left(2Mr-qM^{2}\right)}{r^{2}\Delta}EL_{z}-\frac{\Delta -a^{2}}{r^{2}\Delta}L_{z}^{2}\Bigg]~.
\end{align}
Keeping future utility in mind, we wish to rewrite the above equation in a particular form, by multiplying it throughout by $r^{2}$ and rearranging terms appropriately, yielding, 
\begin{align}
r^{2}\dot{r}^{2}&=-\Delta \delta _{1}+r^{2}E^{2}+\left(a^{2}E^{2}-L_{z}^{2}\right)+\left(\frac{2M}{r}-\frac{qM^{2}}{r^{2}}\right)\left(aE-L_{z}\right)^{2}~.
\label{radgeodold}
\end{align}
Such that, for a massive particle the radial geodesic equation becomes, 
\begin{align}
\dot{r}^{2}&=\frac{1}{r^{4}}\left[-\Delta r^{2}+r^{2}\left(r^{2}+a^{2}\right)E^{2}-r^{2}L_{z}^{2}-\left(-2Mr+qM^{2}+r^{2}+a^{2}\right)\left(aE-L_{z}\right)^{2}
+\left(r^{2}+a^{2}\right)\left(aE-L_{z}\right)^{2}\right]
\nonumber
\\
&=\frac{1}{r^{4}}\left[-\Delta \left\{r^{2}+\left(aE-L_{z}\right)^{2} \right\}+\left(r^{2}+a^{2}\right)^{2}E^{2}+a^{2}L_{z}^{2}
-2aEL_{z}\left(r^{2}+a^{2}\right)\right]
\nonumber
\\
&=\frac{1}{r^{4}}\left[-\Delta \left\{r^{2}+\left(aE-L_{z}\right)^{2} \right\}+\left\{\left(r^{2}+a^{2}\right)E-aL_{z}\right\}^{2}\right]~.
\label{radgeod}
\end{align}
Note that the limit of this equation to the Kerr and the Schwarzschild spacetime is straightforward and yields the desired result for the radial geodesics in these spacetimes. As evident, the effect of the tidal charge arises through the metric coefficient $\Delta(r)$. This equation, along with \ref{brane_E} and \ref{brane_L} describes geodesics moving on the equatorial plane of the rotating braneworld scenario. We would like to emphasize once again that this analysis is going to be unaffected by the nature of the compact objects, i.e., whether the central compact object is a black hole or, an exotic compact object. For simplicity we will assume the above solution to describe a black hole spacetime. 

Having derived the geodesic equations on the equatorial plane, let us concentrate on the determination of the circular geodesics. These geodesics will satisfy both $\dot{r}=0$, as well as $\ddot{r}=0$, yielding two algebraic equations for the conserved energy and angular momentum. Solving these two equations, we obtain the energy and angular momentum associated with a circular orbit on the equatorial plane of the braneworld black hole, located at radius $r$, to yield, 
\begin{align}
E_{(\textrm{circ})}&=\frac{1-\frac{2M}{r}+\frac{qM^{2}}{r^{2}}\pm \frac{a}{r}\sqrt{\frac{M}{r}-\frac{qM^{2}}{r^{2}}}}{\sqrt{1-\frac{3M}{r}+\frac{2qM^{2}}{r^{2}}\pm 2\frac{a}{r}\sqrt{\left(\frac{M}{r}-\frac{qM^{2}}{r^{2}}\right)}}}~,
\label{circ_E}
\\
L_{z(\textrm{circ})}&=\pm \sqrt{Mr}\frac{\left(1+\frac{a^{2}}{r^{2}}\right)\sqrt{1-\frac{qM}{r}}\mp 2\frac{a}{r}\left(\sqrt{\frac{M}{r}}-\frac{qM^{2}}{2r\sqrt{Mr}}\right)}{\sqrt{1-\frac{3M}{r}+\frac{2qM^{2}}{r^{2}}\pm 2\frac{a}{r}\sqrt{\left(\frac{M}{r}-\frac{qM^{2}}{r^{2}}\right)}}}~.
\label{circ_L}
\end{align}
In the above expressions for energy and angular momentum associated with a circular orbit, of radius $r$, on the equatorial plane of the braneworld black hole, the `+' sign corresponds to circular orbits co-rotating with the black hole and `-' sign describes circular orbits counter-rotating to the black hole. The determination of the ISCO requires one additional conditions, namely, not only $\dot{r}$ and $\ddot{r}$ should vanish, but $\dddot{r}$ should vanish as well. These conditions can be understood as follows --- first of all, $\dot{r}=0$ demands that the energy and the effective potential should match at that radius, and then $\ddot{r}=0$ demands that, at this radius the potential should have a minima, such that $(dV/dr)=0$. In general, for stable circular orbits $(d^{2}V/dr^{2})>0$, but the ISCO being the limiting case, actually satisfies $(d^{2}V/dr^{2})=0$ and hence we have to set $\dddot{r}=0$, as well. This yields,
\begin{align}
-2+2E_{(\textrm{circ})}^{2}+\left(\frac{4M}{r^{3}}-\frac{6qM^{2}}{r^{4}}\right)\left(aE_{(\textrm{circ})}-L_{z(\textrm{circ})}\right)^{2}=0~.
\label{isco_rel}
\end{align}
We now have three equations, the expressions for energy and angular momentum follows from \ref{circ_E} and \ref{circ_L}, respectively, while \ref{isco_rel} provides the condition for ISCO. One can substitute for $E_{(\textrm{circ})}$ and $L_{z(\textrm{circ})}$ from \ref{circ_E} and \ref{circ_L}, respectively, in \ref{isco_rel}, and the following algebraic equation satisfied by the radius of the ISCO can be obtained (see \ref{AppA} for a derivation), 
\begin{align}
r^{3}-6Mr^{2}-4q^{2}M^{3}+9qM^{2}r-3a^{2}r+4qMa^{2}\pm 8\left(r-qM\right)a\sqrt{Mr-qM^{2}}=0~.
\label{ISCO}
\end{align}
One can check that in the limit $q\rightarrow 0$, the above algebraic equation reduces to the one satisfied by ISCO in Kerr spacetime. Once again, the `+' sign refers to ISCO, co-rotating with the black hole, while the `-' sign refers to the counter-rotating ISCO, moving opposite to the direction of rotation of the black hole. The solution of the above equation is complicated, and hence will not be presented here in analytic form. We would like to emphasize that the ISCO denotes the transition regime between the in-spiral and the plunge and hence will find numerous use in the subsequent sections. 

\section{Computing fluxes during in-spiral on the brane}\label{sec:FLUXES}

In this section, we will discuss the dynamics of EMRI, in which the central massive black hole inherits contributions from the brane, while the other smaller mass black hole, in-spiraling around the more massive black hole, can be treated as a test particle. The background spacetime due to the massive central black hole has three hairs --- the mass $M$, the rotation parameter $a$, and the contribution from the brane $q$ --- all of these hairs affect the movement of the smaller black hole of mass $m$. Note that, the charge $q$, inherited from the extra dimensions can have both positive and negative signs and we wish to determine how the energy loss in the form of gravitational radiation, arising out of the in-spiral of the smaller mass black hole around the larger mass, gets affected by a non-zero choice for this tidal charge. 

Of course, the general scenario requires a full numerical analysis, however, in this work we will show that analytical results can be derived by ignoring terms $\mathcal{O}(a^{2}/r^{2})$, as well as terms of $\mathcal{O}(aM^{2}/r^{3})$ and $\mathcal{O}(aq/r^{3})$. The above terms can be ignored under the assumptions of slow rotation and large distance approximation, which is valid as long as the smaller mass black hole is at a distance comparable to that of the ISCO of the central massive black hole. The motion of the smaller mass can be derived using the following Lagrangian (for a derivation, see \ref{AppB}), 
\begin{eqnarray}\label{eq:lagrange}
L_{\rm eff}=\frac{m}{2}\Big[\dot{r}^2+r^2 \dot{\theta}^2+r^2 \sin^2\theta \dot{\phi}^2\Big]+\dfrac{m M}{r} -\dfrac{2m a M \sin^2\theta \dot{\phi}}{r}-\dfrac{mqM^2}{2r^2}~,
\label{effL}
\end{eqnarray}
where `dot' denotes derivative with respect to the Boyer-Lindquist time coordinate `t'. It is to be noted that in the $q=0$ limit, it reduces to the corresponding Lagrangian for Kerr spacetime, derived in \cite{Ryan:1995wh}. Also note that for $\theta=(\pi/2)$, i.e., on the equatorial plane, under the small rotation and large distance approximation, the above expression coincides with the one in \ref{brane_L}, with $\dot{t}=1$. Moreover, as evident, the above effective Lagrangian does not depend explicitly on the $\phi$ coordinate and as a consequence, we have the conserved angular momentum $\mathcal{L}_{z}=(\partial L_{\rm eff}/\partial \dot{\phi})$, whose explicit evaluation yields, 
\begin{equation}
\mathcal{L}_{z}=mr^2 \sin^{2}\theta \dot{\phi}-\dfrac{2maM}{r}\sin^2\theta~.
\label{eq:LZ}
\end{equation}
It is possible to invert this relation and obtain, $\dot{\phi}$ in terms of the conserved angular momentum $\mathcal{L}_{z}$. In addition to the angular momentum, due to the existence of a Killing tensor, there is another conserved quantity, namely the Carter constant, which can be expressed as (see \ref{AppB} for a derivation),
\begin{align}\label{carter}
Q+\mathcal{L}_{z}^{2}=m^{2}r^{4}\Big[\dot{\theta}^{2}&+\sin^{2}\theta\dot{\phi}^{2}\Big]
-4am^{2}Mr\sin^{2}\theta\dot{\phi}~.
\end{align}
For the radial coordinate, on the other hand, the geodesic equation derived from the above effective Lagrangian $L_{\rm eff}$ reads,
\begin{eqnarray}
\ddot{r}=r(\dot{\theta}^{2}+\sin^{2}\theta \dot{\phi}^{2})-\dfrac{M}{r^2}+\dfrac{qM^2}{r^3} +\dfrac{2aM\dot{\phi}\sin^{2}\theta}{r^2}~.
\label{eq:rddot}
\end{eqnarray}
Using \ref{eq:LZ}, we can replace the $\sin^{2}\theta \dot{\phi}$ term in terms of the conserved angular momentum $\mathcal{L}_{z}$, in the above geodesic equation for the radial coordinate. Besides, the term involving $\dot{\theta}^{2}$ and $\dot{\phi}^{2}$ can also be expressed in terms of the Carter constant $Q$ and conserved angular momentum $\mathcal{L}_{z}$, from \ref{carter}. Such an exercise yields
\begin{eqnarray}
\ddot{r}=\frac{Q+\mathcal{L}_{z}^{2}}{m^{2}r^{3}}-\dfrac{M}{r^2}+\dfrac{qM^2}{r^3} +\dfrac{6aM\mathcal{L}_{z}}{mr^4}~.
\label{eq:rddot2}
\end{eqnarray}
Therefore, we have arrived at the desired form of the radial geodesic equation. As evident, the above differential equation for the radial coordinate is a second-order equation, with terms involving $r^{-2}$, $r^{-3}$, and $r^{-4}$, respectively. Such a differential equation can be solved exactly in a parametrized form, yielding (see \ref{AppC}), 
\begin{equation}\label{radial}
r=\dfrac{1}{1+e \cos\psi}\left(qM+\frac{Q+\mathcal{L}_{z}^{2}}{m^{2}M}\right)\Bigg[1+\dfrac{6am^{3}M^2\mathcal{L}_{z}}{(Q+\mathcal{L}_{z}^2+qm^{2}M^2)^2}
\left(1+\dfrac{e}{3}\cos\psi\right)\Bigg]~.
\end{equation}
Here $e$ is the eccentricity of the orbit, and we define the parameter $\psi$ as a solution of the differential equation,
\begin{equation}
\dfrac{dt}{d\psi}=\dfrac{\left(Q+\mathcal{L}_{z}^{2}+qm^{2}M^{2}\right)^{3/2}}{m^{3}M^2}\left[1+\dfrac{6am^{3}M^{2}\mathcal{L}_{z}}{\left(\mathcal{L}_{z}^{2}+qm^{2}M^{2}\right)^{2}}\right]\dfrac{1}{(1+e\cos\psi)^2}~.
\end{equation}
Note that the radial coordinate depends explicitly on the tidal charge $q$. For $a=0$, the above solution for $r$ reduces to that of an ellipse, as expected for a Keplerian problem, and $\psi$ in that context would correspond to the angle between the location of the object and its periastron, as seen from the central compact object. 

For equatorial circular orbits, we can set the eccentricity $e=0$ and the Carter constant identically vanishes, so that we have $Q=0$ and the radius of the circular orbit becomes,  
\begin{eqnarray}
r_{\rm circ}=\left(qM+\frac{\mathcal{L}_{z}^{2}}{m^{2}M}\right)\Bigg[1+\dfrac{6am^{3}M^2\mathcal{L}_{z}}{(\mathcal{L}_{z}^2+qm^{2}M^2)^2}\Bigg].
\label{rcirc}
\end{eqnarray}
Interestingly, it is also possible to arrive at the above expression by some approximations. In particular, we can start with the ansatz $r=r_{\rm kerr}+g(q,a)$, where $r_{\rm kerr}$ is the equatorial plane circular orbit with $q=0$, i.e., the Kerr case and $g(q=0,a)=0$. By substituting the above ansatz into \ref{eq:rddot2}, and setting $\ddot{r}=0$, we will arrive at the above expression for the circular orbit, $r_{\rm circ}$ (for a derivation, see \ref{Appd}). 

For non-equatorial but circular orbits, the radius can be derived from \ref{eq:rddot2}, by setting $\ddot{r}=0$ and then solving the corresponding algebraic equation in the radial coordinate $r$, or, by simply setting eccentricity to be zero in \ref{radial}. Either of these yields the same result, i.e.,
\begin{equation}\label{radialcircnoneq}
r_{\rm noneq, circ}=\left(qM+\frac{Q+\mathcal{L}_{z}^{2}}{m^{2}M}\right)\Bigg[1+\dfrac{6am^{3}M^2\mathcal{L}_{z}}{(Q+\mathcal{L}_{z}^2+qm^{2}M^2)^2}\Bigg]~.
\end{equation}
On the other hand, for any circular orbit, the energy associated with the orbit reads (for a derivation, see \ref{energy} in \ref{AppB}).
\begin{align}
\mathcal{E}_{\rm noneq, circ}&=\frac{m}{2}r^{2}\left(\dot{\theta}^{2}+\sin^{2}\theta\dot{\phi}^{2} \right)-\frac{mM}{r}+\frac{mM^{2}q}{2r^{2}},
\nonumber
\\
&=\frac{\left(Q+\mathcal{L}_{z}^{2}\right)+4am^{2}Mr\sin^{2}\theta\dot{\phi}}{2mr^{2}}-\frac{mM}{r}+\frac{mM^{2}q}{2r^{2}},
\nonumber
\\
&=-\frac{mM}{r}+\frac{Q+\mathcal{L}_{z}^{2}+m^{2}M^{2}q}{2mr^{2}}+\frac{2J\mathcal{L}_{z}}{r^{3}},
\end{align}
where, $J=aM$. In the above expression we have used \ref{carter} in order to arrive at the second line, and then we have used \ref{eq:LZ} to get the final expression. Note that the energy $\mathcal{E}$, defined above, differs from the conserved energy $E$, defined in the previous section, by the rest energy of the test particle, aka the smaller mass black hole, i.e., $\mathcal{E}=m(E-1)$. Finally, substituting for the radius of the circular orbit $r$ from \ref{radialcircnoneq}, we obtain the energy of the particle on a non-equatorial circular orbit as, 
\begin{align}
\mathcal{E}_{\rm noneq, circ}&=-\frac{m^{3}M^{2}}{Q+\mathcal{L}_{z}^{2}+m^{2}M^{2}q}\Bigg[1-\dfrac{6m^{3}MJ\mathcal{L}_{z}}{(Q+\mathcal{L}_{z}^2+qm^{2}M^2)^2}\Bigg]+\frac{m^{3}M^{2}}{2\left(Q+\mathcal{L}_{z}^{2}+m^{2}M^{2}q\right)}\Bigg[1-\dfrac{12m^{3}MJ\mathcal{L}_{z}}{(Q+\mathcal{L}_{z}^2+qm^{2}M^2)^2}\Bigg]
\nonumber
\\
&\qquad \qquad +\frac{2m^{6}M^{3}J\mathcal{L}_{z}}{\left(Q+\mathcal{L}_{z}^{2}+m^{2}M^{2}q\right)^{3}}
\nonumber
\\
&=-\frac{m^{3}M^{2}}{2\left(Q+\mathcal{L}_{z}^{2}+m^{2}M^{2}q\right)}\Bigg[1-\dfrac{4m^{3}MJ\mathcal{L}_{z}}{(Q+\mathcal{L}_{z}^2+qm^{2}M^2)^2}\Bigg]~.
\end{align}
It should be emphasized that, even though we mention the above orbit to be circular, it is simply in the sense that $r=\textrm{constant}$, since the orbit is not even confined to a single plane. This can be seen from \ref{eq:LZ} and \ref{carter}, from which it is clear that, 
\begin{align}
\dot{\phi}-\frac{2J}{r^{3}}=\frac{\mathcal{L}_{z}}{r^{2}\sin^{2}\theta}~,
\\
m^{2}r^{4}\left[\dot{\theta}^{2}+\left(\dot{\phi}-\frac{2J}{r^{3}}\right)^{2}\right]=Q+\mathcal{L}_{z}^{2}~,
\label{preOmega}
\end{align}
and hence the effect of rotation is to modify the $\dot{\phi}$ term by $(2J/r^{3})$, with $J=aM$. Thus the motion of the particle is not confined to a plane, rather it precesses about the $x_{3}$ axis with a frequency $(2J/r^{3})$. This information is sufficient to convert the spherical polar coordinates, we have used so far, to the Cartesian coordinate system. This is necessary to obtain the rate of loss of energy due to GW emission during the plunge phase. For that we notice the following properties of the circular orbit --- (a) the circular orbit is in a plane inclined at an angle $\iota$ with the equatorial plane, and (b) the circular orbit precesses about the $x_{3}$ axis with a frequency $(2J/r^{3})$. Therefore, the relation between the spherical polar coordinates and the Cartesian coordinates can be found in the following manner. To start with note that the motion consists of circular motion, such that $x_{1}''=r\cos (\Omega t)$ and $x_{2}''=r\sin (\Omega t)$, with $x_{3}''=0$. Then we perform rotation about the $x_{2}''$ axis by the inclination angle $\iota$, leading to $(x_{1}',x_{2}'',x_{3}')$ and finally, we consider rotation about the modified $x_{3}'$ axis (taken as $x_{3}$, which we take to be the spin axis of the black hole) by the precession angular velocity $(2J/r^{3})$. Therefore, we obtain the following time evolution for the Cartesian coordinates,
\begin{align}
\begin{pmatrix}
x_{1}\\
x_{2}\\
x_{3}
\end{pmatrix}
&=
\begin{pmatrix}
\cos \left(\frac{2Jt}{r^{3}}\right) & -\sin \left(\frac{2Jt}{r^{3}}\right) & 0 \\
\sin \left(\frac{2Jt}{r^{3}}\right) & \cos \left(\frac{2Jt}{r^{3}}\right) & 0 \\
0 & 0 & 1 
\end{pmatrix}
\times 
\begin{pmatrix}
\cos \iota & 0 & -\sin \iota \\
0 & 1 & 0 \\
\sin \iota & 0 & \cos \iota
\end{pmatrix}
\times 
\begin{pmatrix}
r\cos (\Omega t)\\
r\sin(\Omega t)\\
0
\end{pmatrix}
\end{align}
yielding, 
\begin{align}
x_{1}&=r\cos (\Omega t)\cos \left(\frac{2Jt}{r^{3}}\right) \cos \iota-r\sin(\Omega t)\sin \left(\frac{2Jt}{r^{3}}\right)
\\
x_{2}&=r\cos (\Omega t)\sin \left(\frac{2Jt}{r^{3}}\right) \cos \iota+r\sin(\Omega t)\cos \left(\frac{2Jt}{r^{3}}\right)
\\
x_{3}&=r\cos (\Omega t)\sin \iota
\end{align}
where $\Omega$ needs to be determined. Note that $\Omega$ corresponds to the angular velocity of the circular orbit with zero inclination angle, and should be identified with the combination of $\dot{\theta}^{2}$ with $\{\dot{\phi}-(2J/r^{3})\}^{2}$.  With the above expressions, we can now use \ref{preOmega}, and obtain, 
\begin{align}
\Omega&=\sqrt{\dot{\theta}^{2}+\left(\dot{\phi}-\frac{2J}{r^{3}}\right)^{2}}
=\frac{\sqrt{Q+\mathcal{L}_{z}^{2}}}{mr^{2}}
\nonumber
\\
&=\frac{\sqrt{Q+\mathcal{L}_{z}^{2}}}{m\left(qM+\frac{Q+\mathcal{L}_{z}^{2}}{m^{2}M}\right)^{2}}\Bigg[1-\dfrac{12am^{3}M^2\mathcal{L}_{z}}{(Q+\mathcal{L}_{z}^2+qm^{2}M^2)^2}\Bigg]
\nonumber
\\
&=\dfrac{M^2m^3}{(Q+\mathcal{L}^2_{\rm z})^{3/2}}\left(1+\dfrac{qM^2m^2}{Q+\mathcal{L}^2_{\rm z}}\right)^{-2}\Bigg[1-\dfrac{12am^3M^2\mathcal{L}_{\rm z}}{(Q+\mathcal{L}^2_z)^2}\left(1+\dfrac{qM^2m^2}{Q+\mathcal{L}^2_{\rm z}}\right)^{-2}\Bigg]~,
\label{Omega}
\end{align}
where, in the second line we have used the expression for the radius of the circular orbit, from \ref{radialcircnoneq}. Note that in the limit, $q=0$, we get back the angular velocity expression of \cite{Ryan:1995wh}. The same expression for $\Omega$ can also be arrived at from the result that, $\mathcal{L}_{z}=mr^2 \Omega \cos \iota$, and hence using \ref{radialcircnoneq}, along with the result, $\cos \iota=\mathcal{L}_{z}/\sqrt{Q+\mathcal{L}_{z}^{2}}$, we will arrive at \ref{Omega}. 

It is natural to re-write the above expression for the angular velocity $\Omega$ in terms of the velocity of the secondary object moving on a circular trajectory, which we take to be $v^{2}=\{m^{2}M^{2}/(Q+\mathcal{L}_{z}^{2})\}$ \cite{Ryan:1995wh}. In terms of the linear velocity $v$, the angular velocity becomes, 
\begin{align}\label{eq:Omegaisco}
M\Omega=v^{3}\left[1-2qv^{2}-12\left(\frac{a}{M}\right)v^{3}\cos \iota \right]+\mathcal{O}(a^2,qa,q^2)~~.
\end{align}
For $q=0$, it reduces to the corresponding expression for a Kerr spacetime \cite{Ryan:1995wh}, while in the presence of $q$, the angular velocity depends on the tidal charge more strongly than the rotation --- a result, which will be shared by the energy loss through GW as well. 

As emphasized earlier, we need to express all the quantities in Cartesian coordinates, using the simple transformation $x_{1}=r\sin \theta \cos \phi$, $x_{2}=r\sin \theta \sin \phi$, and $x_{3}=r\cos \theta$. In particular, we will be interested in the rate of change of the energy, the angular momentum, and the Carter constant, as the smaller mass black hole makes a transition from one circular orbit to another. Therefore, we should express these conserved quantities in terms of the Cartesian coordinates, leading to, 
\begin{align}
\mathcal{E}&=\frac{m}{2}\dot{x}_{j}\dot{x}_{j}-\frac{mM}{\sqrt{x_{j}x_{j}}}+\frac{mM^{2}q}{2(x_{j}x_{j})}~,
\label{energycart}
\\
\mathcal{L}_{z}&=m\epsilon_{3jk}x_{j}\dot{x}_{k}-\frac{2maM(x_{1}^{2}+x_{2}^{2})}{(x_{j}x_{j})^{3/2}}~,
\label{Lzcart}
\\
Q+\mathcal{L}_{z}^{2}&=m^{2}\left(\epsilon_{ijk}x_{j}\dot{x}_{k}\right)\left(\epsilon_{ilm}x_{l}\dot{x}_{m}\right)-\frac{4am^{2}M \epsilon_{3jk}x_{j}\dot{x}_{k}}{\sqrt{x_{j}x_{j}}}~.
\label{cartercart}
\end{align}
In order to reconcile the above expressions with their spherical polar counterparts, we note the following results, $\epsilon_{3jk}x_{j}\dot{x}_{k}=x_{1}\dot{x}_{2}-x_{2}\dot{x}_{1}=r\sin \theta \cos \phi(\sin \theta \sin \phi \dot{r}+r\cos \theta \sin \phi \dot{\theta}+r\sin \theta \cos \phi \dot{\phi})-r\sin \theta \sin \phi(\sin \theta \cos \phi \dot{r}+r\cos \theta \cos \phi \dot{\theta}-r\sin \theta \sin \phi \dot{\phi})=r^{2}\sin^{2}\theta \dot{\phi}$. Also, $(\epsilon_{ijk}x_{j}\dot{x}_{k})(\epsilon_{ilm}x_{l}\dot{x}_{m})=(x_{j}x_{j})(\dot{x}_{k}\dot{x}_{k})-(x_{j}\dot{x}_{j})^{2}=r^{2}\times (\dot{r}^{2}+r^{2}\dot{\theta}^{2}+r^{2}\sin^{2}\theta \dot{\phi}^{2})-r^{2}\dot{r}^{2}=r^{4}(\dot{\theta}^{2}+\sin^{2}\theta \dot{\phi}^{2})$. Note that besides energy, neither $\mathcal{L}_{z}$, nor $Q+\mathcal{L}_{z}^{2}$ depend upon the presence of the tidal charge $q$ explicitly, and hence \ref{Lzcart} and \ref{cartercart} coincides with the expressions in \cite{Ryan:1995zm}. 

As we have indicated before, for the case of our interest, e.g., the EMRI system, the above constants of motion are not really constants, since the lower mass compact object inspiralling around the more massive compact object will slowly move towards smaller and smaller radii. Therefore, the above constants will change due to the radiation reaction from the emitted GW. These changes can be quantified by taking derivatives of \ref{energycart} to \ref{cartercart}, with respect to time. These yields, in the Cartesian coordinates,
\begin{align}
\dot{\mathcal{E}}&=m\dot{x}_{j}\ddot{x}_{j}+\frac{mM(x_{j}\dot{x}_{j})}{(x_{j}x_{j})^{3/2}}
-\frac{mM^{2}q(x_{j}\dot{x}_{j})}{(x_{j}x_{j})^{2}}~,
\label{energycartdot}
\\
\dot{\mathcal{L}}_{z}&=m\epsilon_{3jk}x_{j}\ddot{x}_{k}-\frac{4maM(x_{1}\dot{x}_{1}+x_{2}\dot{x}_{2})}{(x_{j}x_{j})^{3/2}}+\frac{6maM(x_{1}^{2}+x_{2}^{2})(x_{j}\dot{x}_{j})}{(x_{j}x_{j})^{5/2}}~,
\label{Lzcartdot}
\\
\dot{Q}+2\mathcal{L}_{z}\dot{\mathcal{L}}_{z}&=2m^{2}\left(\epsilon_{ijk}x_{j}\dot{x}_{k}\right)\left(\epsilon_{ilm}x_{l}\ddot{x}_{m}\right)-\frac{4am^{2}M \epsilon_{3jk}x_{j}\ddot{x}_{k}}{\sqrt{x_{j}x_{j}}}
+\frac{4am^{2}M \epsilon_{3jk}x_{j}\dot{x}_{k}(x_{m}\dot{x}_{m})}{(x_{j}x_{j})^{3/2}}~.
\label{cartercartdot}
\end{align}
In the above expressions, we are only interested in the radiation reaction contributions to the rate of change of these quantities, and hence the term $\ddot{x}_j$, which we refer to as $a_j$ is the one of importance. Therefore, apart from the first term in the energy and the angular momentum expressions, and the first and second terms in the Carter constant expression, we will not consider the other terms. It is to be emphasized that we are only taking the terms which have an acceleration component since our interest is in the radiative part, and we are \emph{not} ignoring any terms in the above expressions. This will lead to the following results for the radiative change of the conserved quantities,
\begin{align}
\dot{\mathcal{E}}^{({\rm rad})}=m\dot{x}_{j}a_{j}~, 
\quad 
\dot{\mathcal{L}}_{z}^{({\rm rad})}=m\epsilon_{3jk}x_{j}a_{k}~, 
\quad 
\dot{\left(Q+\mathcal{L}_{z}^{2}\right)}^{({\rm rad})} =2m^{2}\left(\epsilon_{ijk}x_{j}\dot{x}_{k}\right)\left(\epsilon_{ilm}x_{l}a_{m}\right)-\frac{4am^{2}M \epsilon_{3jk}x_{j}a_{k}}{\sqrt{x_{j}x_{j}}}~.
\label{alldot}
\end{align}
For the acceleration term $a_j$, arising out of radiation reaction forces, we follow the prescription given in Ref. \cite{Ryan:1995zm} which in the Cartesian coordinates read,
\begin{eqnarray}
a_j=-\dfrac{2}{5}I^{(5)}_{jk}x_{k}+\dfrac{16}{45}\epsilon_{jpq}J^{(6)}_{pk}x_{q}x_{k}+\dfrac{32}{45}\epsilon_{jpq}J^{(5)}_{pk}x_{k} \dot{x}_{q}+\dfrac{32}{45}\epsilon_{pq[j}J^{(5)}_{k]p}x_{q} \dot{x}_{k}+\dfrac{8J}{15}J^{(5)}_{3i}.
\end{eqnarray}
In the above, we have introduced the anti-symmetry notation, i.e., $A_{[ij]}=(1/2)(A_{ij}-A_{ji})$, and the superscripts in the brackets denote how many times the respective quantity has been differentiated with respect to time, e.g., $I^{(5)}_{jk}=(d^{5}/dt^{5})I_{jk}$. Moreover, the quantities $I_{jk}$ and $J_{jk}$ has the following expressions in the Cartesian coordinates,
\begin{equation}
I_{ij}=\Big[mx_i x_j\Big]^{\text{STF}}~;
\qquad
J_{ij}=\Big[mx_{i}\epsilon_{jkm}x_{k}\dot{x}_{m}-\dfrac{3}{2}mx_iJ\delta_{j3}\Big]^{\text{STF}}~,
\end{equation}
with \enquote{STF} standing for the symmetric and trace-free part of the tensor. Note that we are only keeping terms involving mass and current quadrupole moments, this is because, all the higher order moments of mass and current multipoles involving terms $\mathcal{O}(a^{2})$, $\mathcal{O}(aq)$ and $\mathcal{O}(q^{2})$, respectively, have been ignored. To express the rate of change of energy, angular momentum, and Carter constants in terms of the parameters of the problem, e.g., the mass, the rotation parameter, and the extra-dimensional charge, we need to substitute the above expression for acceleration in \ref{alldot}. Subsequently, we have to average out over a complete period of revolution. With this approach, we obtain the following expression for the energy flux:
\begin{equation}
\langle \dot{\mathcal{E}}\rangle^{({\rm rad})}=-\dfrac{32}{5}\eta^2v^{10}\Big[1-8qv^2-\dfrac{433}{12}(a/M)v^3\cos \iota\Big]+\mathcal{O}(a^2,qa,q^2),
\label{eq:DEDT}
\end{equation}
where, $\eta$ is the mass ratio, defined as $m/M$, with $m$ being the mass of the smaller object and $M$ is the mass of the heavier object. Similarly, we can also construct the flux due to angular momentum and Carter constant, averaged over a complete period of revolution, yielding,
\begin{eqnarray}
\langle \dot{\mathcal{L}}_{\rm z}\rangle^{({\rm rad})}&=&-\dfrac{32}{5}\eta^2v^7 \Big[(1-6qv^2)\cos \iota+\dfrac{61-687 \cos^2\iota}{24}(a/M)v^3\Big]+\mathcal{O}(a^2,qa,q^2)~,
\label{eq:DLDT}
\\
\langle\dot{Q+\mathcal{L}^2_{\rm z}}\rangle^{({\rm rad})}&=& -\dfrac{64}{5}\eta^2 v^6 \Big[1-6 qv^2-\dfrac{313}{12}(a/M)v^3  \cos \iota\Big]+\mathcal{O}(a^2,qa,q^2)~.
\label{eq:DQDT}
\end{eqnarray}
The above flux laws show an intriguing feature, the effect of the tidal charge appears at a lower post-Newtonian (PN) order than that of the rotation. The leading term in the energy flux balance law corresponds to a 2.5 PN term, while the effect of charge appears at 3.5 PN and that of rotation at 4 PN. Therefore, the presence of a small tidal charge, originating from the extra dimension, can affect the energy loss formula by a considerable amount, simply because, it is appearing at a lower PN order. For example, if we take the inclination angle to be $\iota \sim 77^{\rm o}$, then it follows that for the ratio $(qM/a)\sim v$, the contributions from both the tidal charge and the rotation will be of the same order. This implies that the existence of a tidal charge can indeed contribute significantly to the GW energy loss. We will observe similar conclusions from the subsequent sections as well. This is also consistent with similar explorations with charge studied in various contexts \cite{Mukherjee:2020how}. In the limiting case of $q=0$, the above relations match with the expressions given in Refs. \cite{Ryan:1995zm,Ryan:1995xi}. 

\section{Modeling the inspiral to plunge phase in braneworld gravity}\label{transitionbrane}

In this section, we will discuss the transition regime from the inspiral to the plunging phase of an EMRI in the braneworld scenario. We closely follow the seminal work by Ori and Thorne in Ref. \cite{Ori:2000zn}, which describes this phase in the Einstein's gravity. By using the aforementioned fluxes in \ref{sec:FLUXES}, we evolve the system close to ISCO and study how the presence of the tidal charge $q$ is affecting the inspiral. In addition to that, we briefly discuss an elementary analysis to obtain the accumulated SNR during this phase, and investigate whether $q$ can introduce considerable change in the GW signal in order to render it detectable. 

\subsection{Evolving braneworld EMRI with fluxes}\label{braneemriflux}

In our simple setup, the secondary, i.e., the less massive object is moving on a circular orbit around the primary, i.e., the more massive central object. The secondary object is considered as a point particle, while the spacetime around the primary object contains the signature of the higher dimensional spacetime, in the form of the tidal charge $q$, introduced in the previous section. To start with, we note that the rate of change of energy carried out by the GW is simply the negative of the rate of change of energy radiated by the system, as given by \ref{eq:DEDT}. This in turn results into the orbit to shrink, at a rate, given by the following expression, 
\begin{align}
\frac{dr}{dt}=\frac{\dot{\mathcal{E}}}{\left(d\mathcal{E}/dr\right)}=\frac{\dot{E}}{(dE/dr)}~.
\end{align}
Here, we have used the result $\mathcal{E}=m(E-1)$, where $E$ is the relativistic energy per unit mass, such that $\dot{E}=m\dot{E}$ and $(d\mathcal{E}/dr)=m(dE/dr)$. Since, $\dot{\mathcal{E}}$ and hence $\dot{E}$ is a negative quantity, as evident from \ref{eq:DEDT}, it follows that $(dr/dt)$ is negative as well, implying that the secondary object moves closer to the primary object as time increases, consistent with our expectations. Note that the time coordinate used here corresponds to the Boyer-Lindquist time coordinate. 

As the secondary object inspirals towards the primary, it does so adiabatically in the inspiral phase and then as it nears the ISCO, the transition regime to the final plunge into the primary object begins, and the evolution of the motion ceases to be adiabatic. However, till the final plunge scenario is reached, i.e., during the transition regime, which is close to the innermost stable circular orbit, and evolution is still primarily due to the radiation reaction from the outgoing gravitational radiation. After reaching the plunge phase, the effect of the radiation reaction of the GWs on the evolution of the orbit can be ignored. Since the transition regime happens close to the ISCO, the energy and the angular momentum of the secondary object can be taken to be constant and equal to their values for the ISCO. These values are simply given by \ref{circ_E} and \ref{circ_L}, with the radius being equal to that ISCO, which is a solution to \ref{isco_rel}. Since the transition phase happens close to the ISCO, for our purpose it will be sufficient to write down the following relation between energy per unit secondary mass $E$ and the angular momentum per unit secondary mass $L_{z}$ as, 
\begin{equation}\label{ELisco}
E=E_{\rm isco}+\Omega_{\rm isco}\Gamma~; 
\qquad 
\Gamma\equiv L_{z}-L_{z,\,\textrm{isco}}~,
\end{equation}
where, the subscript \enquote*{isco} denotes the relevant quantities for the ISCO. Notice that the above equation can also be cast in terms of the following quantities, $\widetilde{L}_{\rm z}=L_{\rm z}/M$ and $\widetilde{\Omega}_{\rm isco}=M\Omega_{\rm isco}$ and has an identical structure. The above relation arises from the assumption that close to the ISCO, the rate of change of energy and the rate of change of angular momentum are related through the angular velocity at the ISCO, such that,
\begin{equation}
\dfrac{dE}{d\tau}=\Omega_{\rm isco} \dfrac{d\Gamma}{d\tau}~,
\label{dEOdL}
\end{equation}
where, $\tau$ is the proper time associated with the trajectory for the secondary object. The implication being, by computing either one among the flux of energy and angular momentum, we can obtain the other one. The above relation can be further modified by changing the derivative of energy with respect to the proper time to a derivative with respect to the Boyer-Lindquist time coordinate $t$, such that,
\begin{eqnarray}
M\dfrac{dE}{d\tau}=\frac{M}{m}\dfrac{d\mathcal{E}}{d\tau}=\dfrac{1}{\eta}\left(\dfrac{dt}{d\tau}\right)\dfrac{d\mathcal{E}}{dt}~,
\end{eqnarray}
Therefore, substituting the above equation for $(dE/d\tau)$ in \ref{dEOdL}, we can finally write
\begin{eqnarray}
M\dfrac{d\Gamma}{d\tau}=\dfrac{M}{\Omega_{\rm isco}}\dfrac{dE}{d\tau}=\dfrac{1}{\eta\Omega_{\rm isco}}\left(\dfrac{dt}{d\tau}\right)\dfrac{d\mathcal{E}}{dt}~, 
\label{eq:dxdtau}
\end{eqnarray}
Since we are working with circular and equatorial orbits, the average over a complete revolution is the same as the instantaneous expression for $d\mathcal{E}/dt$, and hence by substituting the expression of the radiative energy flux $\dot{\mathcal{E}}^{({\rm rad})}$ from \ref{eq:DEDT}, we arrive at:
\begin{equation}\label{eq:Radiation_EN}
\dfrac{d\Gamma}{d\tau}=-k \eta~,   
\end{equation}
with 
\begin{equation}
k\equiv \dfrac{32}{5}\left(\dfrac{dt}{d\tau}\right)\left(\frac{1}{M\Omega_{\rm isco}}\right)v^{10} \Big[1-8qv^2-\dfrac{433}{12}\left(\frac{a}{M}\right)v^{3}\Big]~.
\end{equation}
As our interests are in the equatorial orbits, we have vanishing Carter constant $Q$, and hence the linear velocity of the secondary object simply becomes, $v=(M/L_{z})$. The benefit of writing down the rate of change of angular momentum per unit mass as above highlights certain points --- firstly, the explicit mass dependence of \ref{eq:Radiation_EN} comes through $\eta$, since the quantity $k$ is dimensionless. This implies that the momentum/energy is decaying very slowly for an EMRI system, as $\eta$ is a small number for the case of extreme mass ratio. Secondly, for a given value of the black hole's parameters, $k$ is a constant and depends on the properties of the ISCO. Therefore, \ref{eq:Radiation_EN} can be immediately integrated, with the boundary condition, $\Gamma(\tau=0)=0$, i.e., the transition regime starts at the ISCO, to obtain,
\begin{eqnarray}\label{Gammaexp}
\Gamma=-k\eta \tau~. 
\end{eqnarray}
Thirdly, presence of the tidal charge $q$ and most importantly, its sign will further affect the decay rate. Therefore, a true braneworld scenario, involving a negative tidal charge can be distinguished from the existence of a Maxwell field through the rate of change of angular momentum, since the rate depends explicitly on the sign of $q$.

\begin{figure}[htp]
\centering
\includegraphics[scale=0.35]{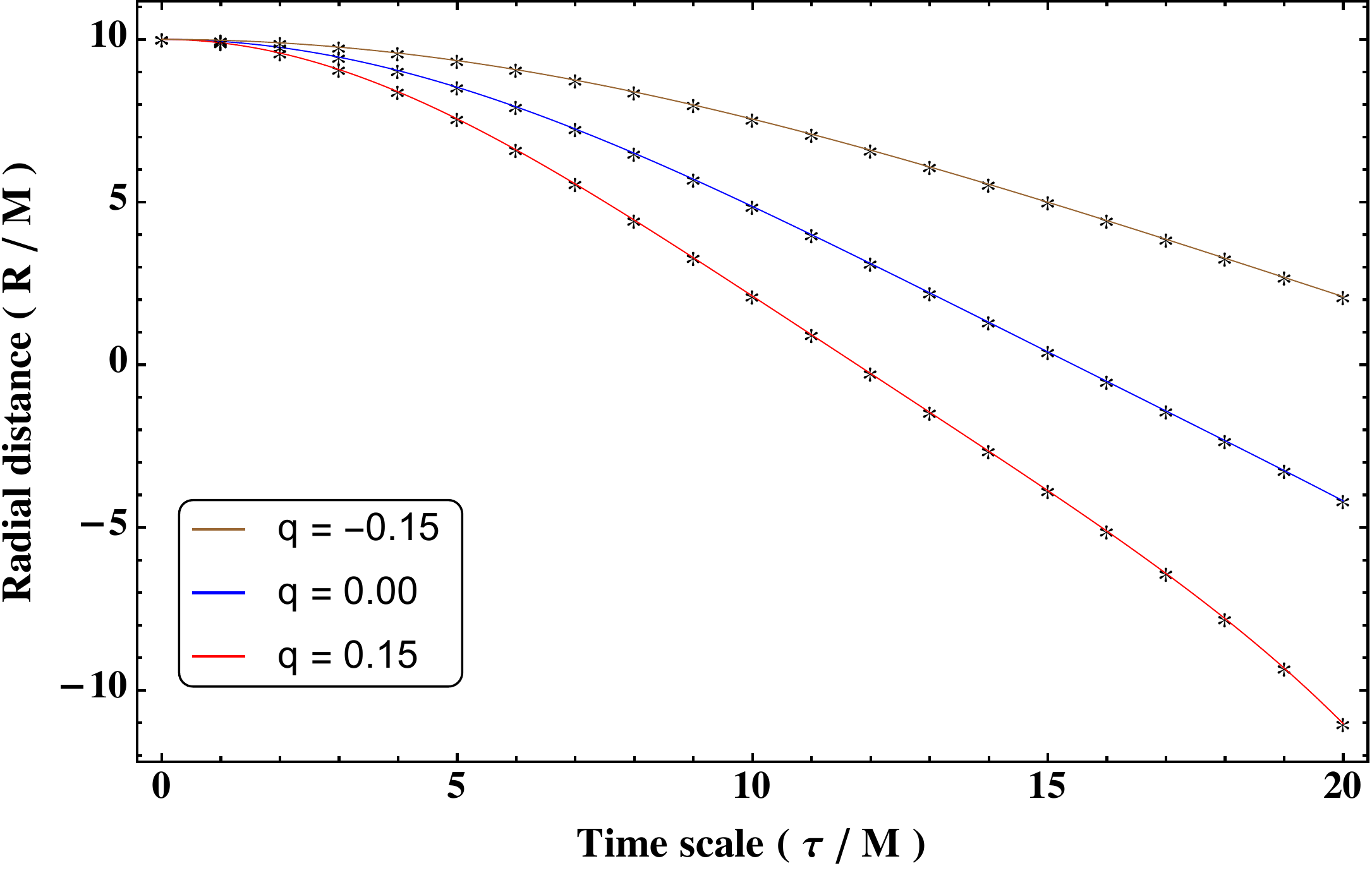}
\includegraphics[scale=0.34]{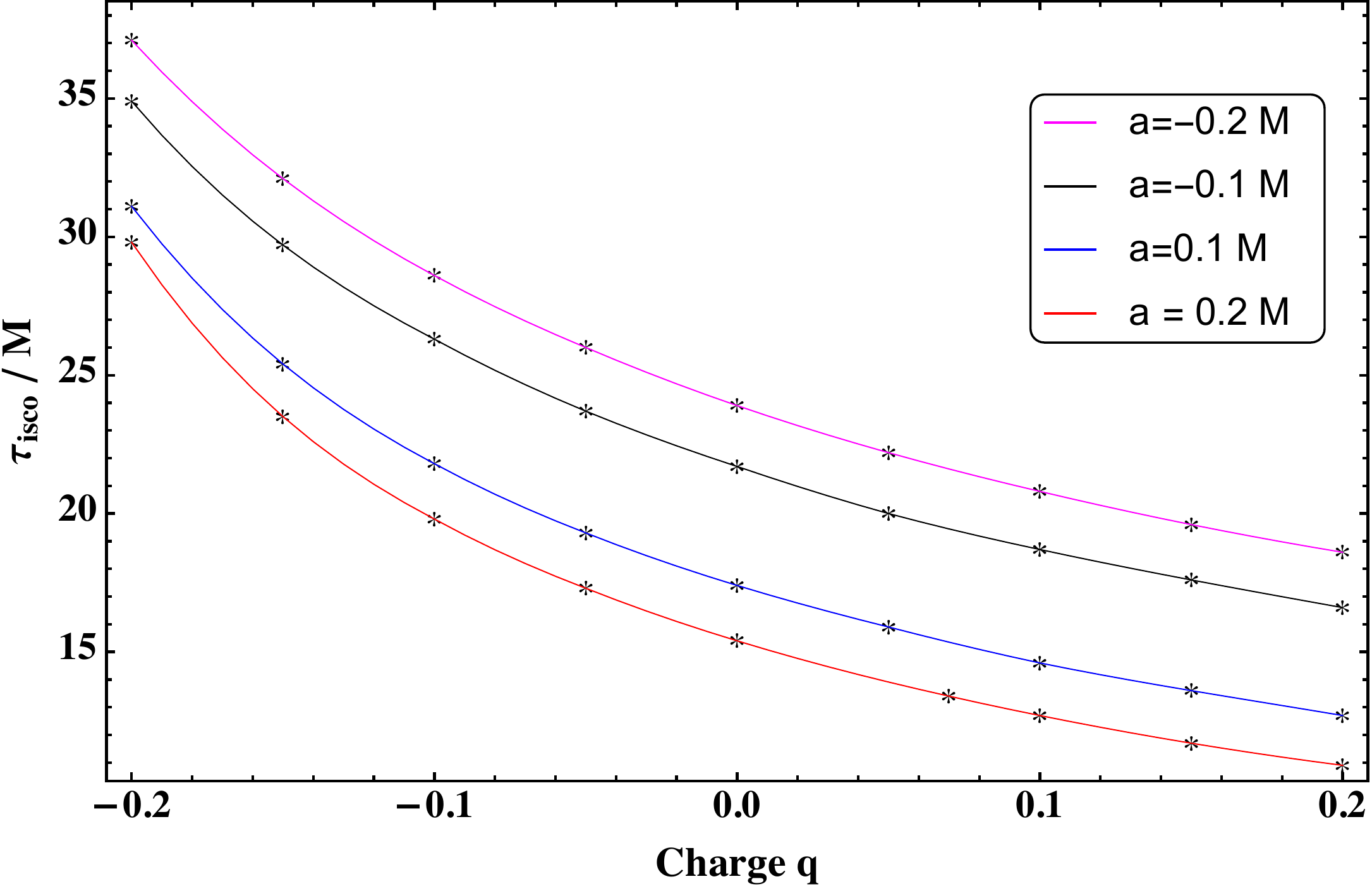}
\caption{The above figures demonstrate how the radial coordinate in the transition regime changes as a function of the proper time for various values of the tidal charge $q$ in the braneworld parameter and the rotation parameter $a$. For the left figure, we set the rotation parameter of the BH to be $a=0.1M$ and it depicts that the radial coordinate changes very slowly for negative values of the tidal charge $q$ compared to its positive values. Here, $\tau=0$ is the time slice, from where the plunge phase starts. Note that the starting of the plunge phase is somewhat arbitrary, but should be around the ISCO radius. In the right figure, $\tau_{\rm isco}$ corresponds to the time taken by the secondary object to reach the ISCO radius, and we have plotted the same as a function of the tidal charge $q$, for different values of the black hole rotation parameter $a$. Note that, on the right-hand figure, we have presented both co-rotating and counter-rotating configurations. See text for further discussions.}
\label{fig:figure_01}
\end{figure}

Given the above expressions for the rate of change of energy and angular momentum, we can also obtain the change of the radial coordinate $r$ as the EMRI evolves from the ISCO, through the transition regime, towards the final plunge phase. For this purpose, we start with the geodesic equation along the radial direction, 
\begin{equation}
\left(\dfrac{dr}{d\tau}\right)^2=E^2-V_{\rm eff}(r,E,L_{z}),
\end{equation}
which is an alternative form to \ref{radgeod} and $V_{\rm eff}(r,E,L_{z})$ is the effective potential, having the following expression,
\begin{eqnarray}
V_{\rm eff}(r,E,L_{z})=E^2-\dfrac{1}{r^4}\left[\left\{E\left(r^2+a^2\right)-aL_z\right\}^2-\left(r^2-2Mr+a^2+qM^{2}\right)\left\{r^2+(L_z-aE)^{2}\right\}\right]~.
\end{eqnarray}
Note that $E$ being the energy per unit mass, is dimensionless and $L_{z}$ has the dimension of length, such that the effective potential is dimensionless. Moreover, from our previous analysis it is clear that at ISCO, we have $V_{\rm eff}(r_{\rm isco})=E_{\rm isco}^{2}$, as well as $(\partial V_{\rm eff}/\partial r)$ and $(\partial^{2}V_{\rm eff}/\partial r^{2})$ must vanish at the ISCO radius. Using these and expanding the energy and the angular momentum as well about their values at the ISCO, we obtain,
\begin{eqnarray}
V_{\rm eff}(r,E,L_{z})=E_{\rm isco}^{2}+\frac{1}{3!}\left(\frac{\partial^{3}V_{\rm eff}}{\partial r^{3}}\right)_{\rm isco}R^{3}
+\left(\frac{\partial V_{\rm eff}}{\partial L_{z}}+\frac{\partial V_{\rm eff}}{\partial E}\Omega\right)_{\rm isco}\Gamma
+\frac{1}{2}\left(\frac{\partial^{2} V_{\rm eff}}{\partial L_{z}\partial r}+\frac{\partial^{2} V_{\rm eff}}{\partial E\partial r}\Omega\right)_{\rm isco}\Gamma R+\mathcal{O}(\Gamma^2),\nonumber \\
\end{eqnarray}
where, we have defined a new radial coordinate $R \equiv r-r_{\rm isco}$ and have kept only leading order terms in the radial coordinate $R$ and the angular momentum difference $\Gamma$. In addition, we have used \ref{ELisco} to replace the energy in terms of $\Gamma$. For brevity, we rewrite the above effective potential as,
\begin{equation}
V_{\rm eff}(R,\Gamma)=\dfrac{2\alpha}{3}R^3-2 \beta \Gamma R+\gamma \Gamma+\text{constant}~,
\end{equation}
and hence the evolution of the newly defined radial coordinate $R$ through the transition regime can be obtained by solving the following differential equation, 
\begin{equation}
\dfrac{d^2R}{d\tau^2}=-\dfrac{1}{2}\dfrac{\partial V_{\rm eff}(R,\Gamma)}{\partial R}=-\alpha R^2+\beta  \Gamma=-\alpha R^2-\eta \beta k \tau~.
\label{eq:R_evolve}
\end{equation}
We have used \ref{Gammaexp} in order to arrive at the final expression for the evolution of the radial coordinate $R$. Moreover, the expressions for the constants $\alpha$ and $\beta$, appearing in the expression for the effective potential are given as,
\begin{eqnarray}
\alpha&=&\dfrac{1}{4}\left(\dfrac{\partial^3 V}{\partial r^{3}}\right)_{\rm isco}
=3\left[\frac{M}{r^4}+\frac{2}{r^5}\Big\{a^2(-1+E^2)-L^2_z-q M^2\Big\}+10\frac{M}{r^{6}} (L_z-aE)^2-10 (L_z-aE)^2 \frac{qM^2}{r^{7}}\right]_{\rm isco}~,  
\\
\beta&=&-\dfrac{1}{4}\Big(\dfrac{\partial^2 V}{\partial L_{\rm z} \partial r}+\Omega\dfrac{\partial^2 V}{\partial E\partial r}\Big)_{\rm isco}
=2\Big[\frac{L_z-a^2 E\Omega}{r^3}-3\frac{M}{r^{4}}(L_z-aE)(1-a \Omega)+2\frac{qM^2}{r^{5}}(1-a\Omega)(L_z-aE)\Big]_{\rm isco}~. 
\end{eqnarray}
Note that both $\alpha$ and $\beta$ are dimension-full quantities, and can be converted to a dimensionless one, by rescaling the radial coordinate $R$, the angular momentum $L_{z}$ and the rotation parameter $a$ by the mass $M$, while the angular velocity needs to be rescaled by the inverse of the mass. With the above expressions, we can now attempt to evaluate how the presence of the tidal charge $q$ is affecting the transition from the inspiral to the plunge. The differential equation for the evolution of the radial coordinate cannot be solved analytically and hence we solve it using numerical routines for solving differential equations. The results of which have been presented in \ref{fig:figure_01}. The left-hand figure, in \ref{fig:figure_01}, depicts the evolution of the radial distance from the ISCO, during the transition regime, with the proper time of the secondary object. As evident, the evolution of the radial coordinate is different for different values and signs of the tidal charge. In particular, for a negative tidal charge, the rate of change of the radial coordinate is slow, and the secondary object takes a longer time to reach the plunge phase. By changing the sign of the charge to be positive, we observe that it has similar effects to that of the black hole's spin -- they both accelerate the inspiral phase, and the plunge phase is reached more quickly. As one approaches radii far beyond ISCO, the adiabatic approximation is likely to deviate strongly from the true inspiral and ultimately lead to divergent results. This behavior is visible from \ref{fig:figure_01} and can be seen more explicitly if we extend $\tau$ to larger values. 
The right hand figure in \ref{fig:figure_01} depicts how the proper time taken to reach the ISCO changes with the tidal charge $q$ and the rotation parameter $a$. As evident, for a given tidal charge $q$, increasing the rotation parameter, leads to a shorter proper time to reach ISCO. On the other hand, for a given rotation parameter, if the tidal charge is negative, it takes a longer time to reach ISCO, while for a positive tidal charge, it takes shorter time to reach ISCO. These are consistent with our earlier findings as well. Having derived the general properties of the transition phase and its nature of dependence on the tidal charge and rotation parameter, we now turn to the observational consequences of the tidal charge and its detection is the future GW experiments.

\subsection{Possible observational implications}\label{braneemriobsrv}

\begin{figure}[htp]
\centering
\includegraphics[scale=0.35]{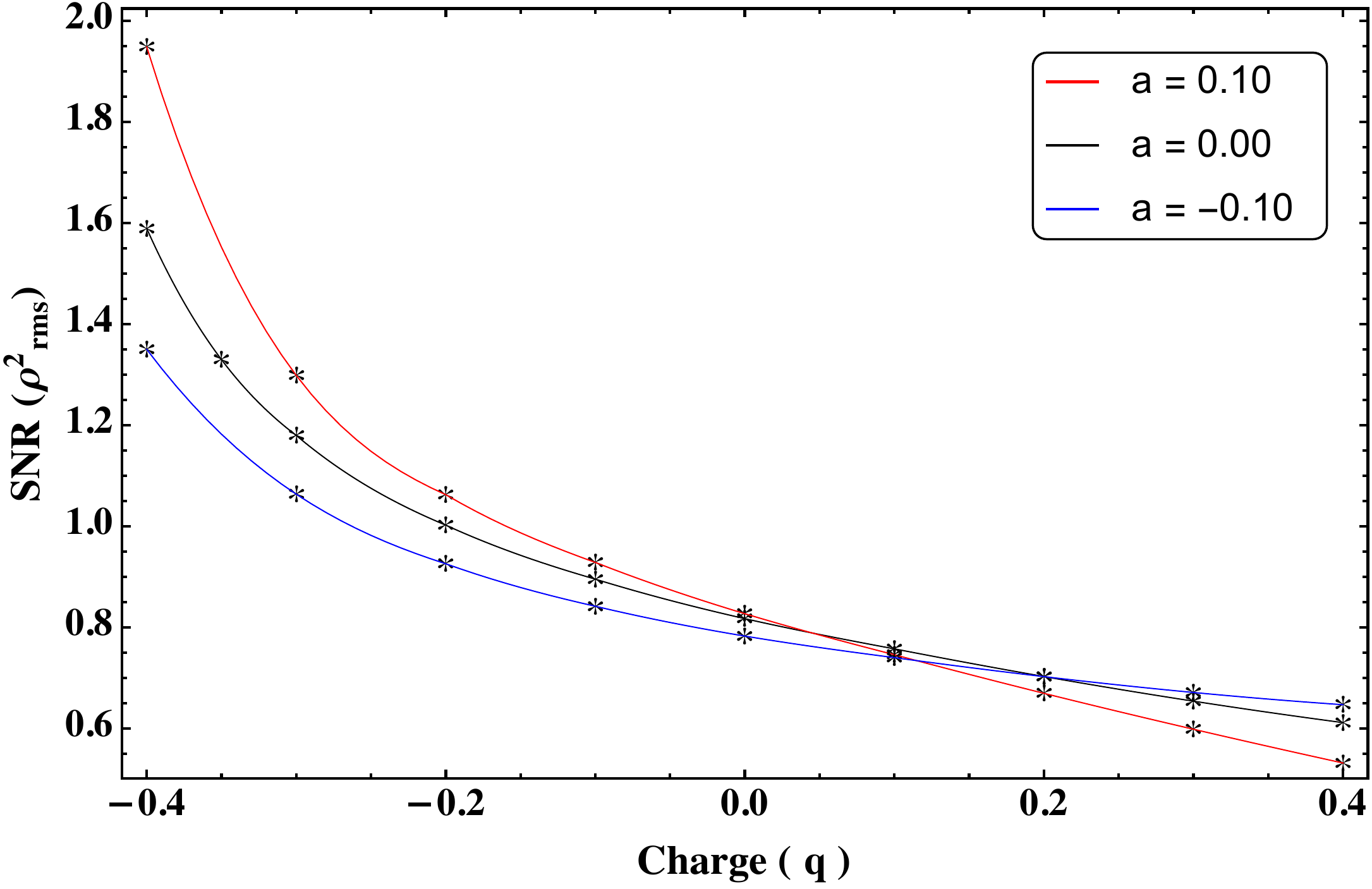}
\includegraphics[scale=0.35]{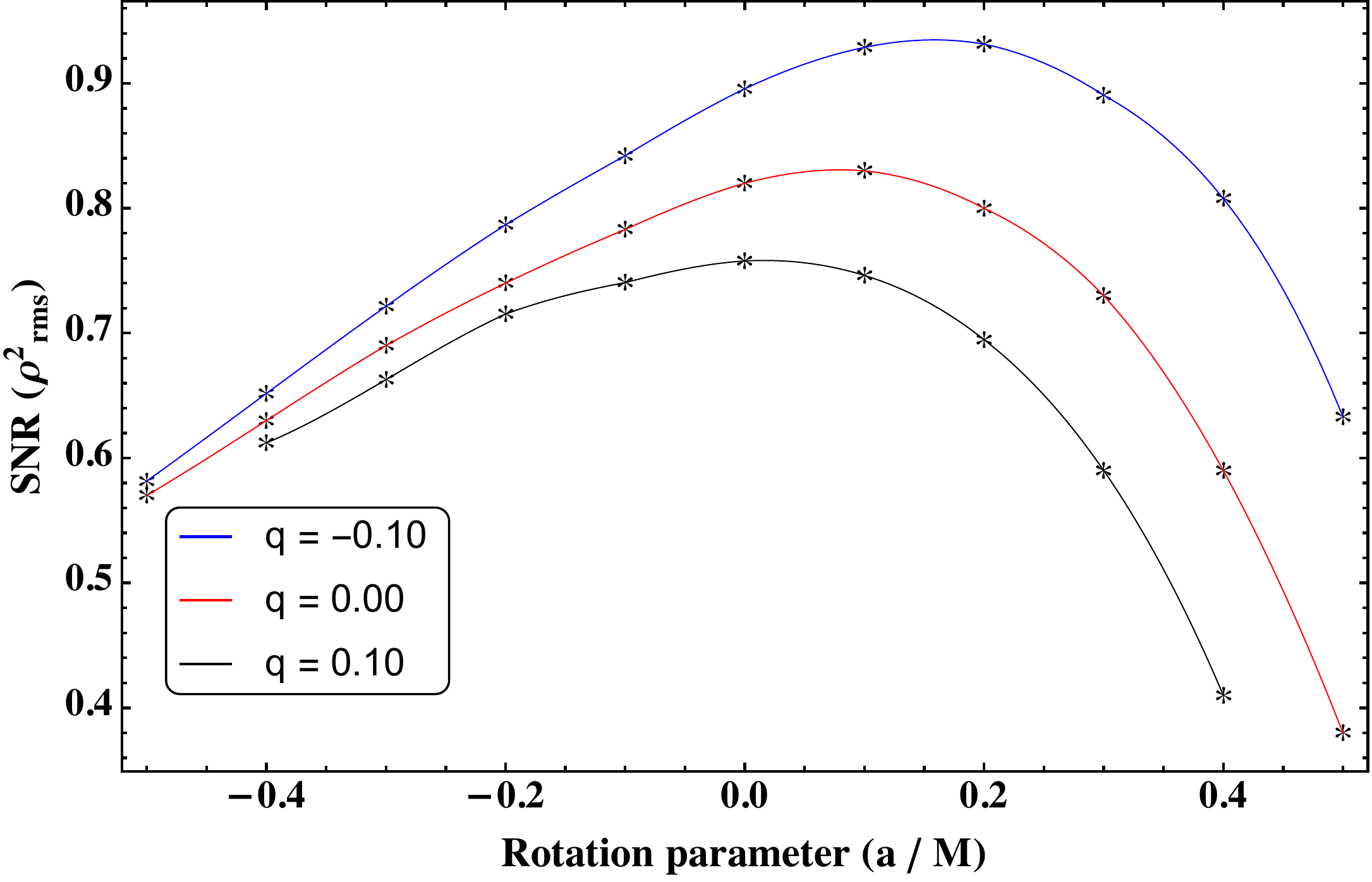}
\caption{In the above figures, we depict the SNR values as a function of the tidal charge $q$ and the rotation parameter $a$. We have used \ref{eq:rhorms} with mass ratio $\eta=10^{-5}$, distance $D=1~\text{Gpc}$ \cite{Ori:2000zn}. The value of $\Delta t$ is different for different orbital parameters. This is given in \ref{Tab_01} and \ref{Tab_02}. We have presented the SNR values for both the co-rotating ($a>0$) and the counter-rotating ($a<0$) orbital configurations. In the left figure, the SNR is shown for different values of the rotation parameter, as a function of the charge. For the negative values of the tidal charge, the SNR is much higher compared to the positive values. Also, for negative $q$, a larger rotation gives larger SNR. This feature is flipped in the positive branch, i.e., for $q>0$. Interestingly enough, there is a peak in the right figure, which shows that positive values of the rotation parameter $a$, i.e., when the motion during the transition regime from inspiral to plunge happens along the direction of rotation of the primary BH spacetime, is going to be dominant in the GW signal. Moreover, negative values of the tidal charge enhance the SNR. Thus, the existence of an extra spatial dimension will have a distinct signature on the plunge phase of the GW waveform of an extreme mass ratio inspiral.}
\label{fig:figure_02}
\end{figure}

During the transition from the inspiral to the plunge, the binary system radiates energy and angular momentum in the form of GWs. Given that our set up is based on the assumption of circular and equatorial geodesic, the frequency of this GW, denoted as $\Omega_{\rm GW}$, is twice of the orbit's angular velocity. In our case, the only angular velocity of importance is that of the ISCO, and therefore, $\Omega_{\rm GW}=2\Omega_{\rm isco}$. Let us now consider that the radiated energy is uniformly distributed, i.e., isotropic, and therefore, with per unit time $dt$, and per unit area $dA$, we have
\begin{equation}
\dfrac{d\mathcal{E}}{dAdt}= \dfrac{\dot{\mathcal{E}}}{4\pi D^2},
\end{equation}
where $D$ is the distance between the source and the detector. Note that since we are discussing about EMRI systems, the detector will be LISA, since that is sensitive to the signals with frequency in the mHz range. Following Ref. \cite{Maggiore:2007ulw}, we can relate $\dot{\mathcal{E}}$ with the plus ($h_{+}$) and cross ($h_{\times}$) polarization as follows: 
\begin{equation}
\dfrac{\dot{\mathcal{E}}}{4\pi D^2}=\dfrac{\langle{\dot{h}^2_{\rm +}+\dot{h}^2_{\rm \times}\rangle_{\rm avg}}}{16\pi}.
\label{eq:EDOT_h}
\end{equation}
where, we have to perform the sky average of the GW perturbations. Note that the expression for $\dot{\mathcal{E}}$ in the braneworld scenario has already been derived and appears in \ref{eq:DEDT}. Following which, and assuming that GWs satisfy a plane wave solution, we have $h \sim \exp(i\Omega_{\rm GW}t)$ for both the polarizations, the sky average can be calculated and finally we obtain,
\begin{equation}\label{hrms}
h^2_{\rm rms}=\langle{h^{2}_{+}+h^2_{\times}\rangle}_{\rm avg}=\dfrac{8\dot{\mathcal{E}}}{ D^2}\dfrac{1}{(2\pi f)^2}=\dfrac{8\dot{\mathcal{E}}}{ D^2}\dfrac{1}{\Omega_{\rm GW}^2}=\dfrac{2\dot{\mathcal{E}}}{ D^2}\dfrac{1}{\Omega_{\rm isco}^2}~.
\end{equation}
where, we have used the result that the frequency of the GWs are twice of the angular velocities of the secondary object around the primary, generating the GWs. With the above expression for $h_{\rm rms}$ being obtained, we may introduce the noise spectral density of LISA, namely $S_{\rm h}(f)$, which takes the form \cite{Robson:2018ifk},
\begin{equation}
S_{h}(f)=\dfrac{10}{3L^2}\Big[P_{\rm OMS}(f)+2\left\{1+\cos^2(f/f_{*})\right\}\dfrac{P_{\rm acc}}{(2\pi f)^4}\Big]\Big[1+\dfrac{6}{10}\Big(\dfrac{f}{f_{*}}\Big)^2\Big]~,
\end{equation}
where $L$ is the arm-length of the detector and $f_{*}$ is the frequency at which the detector has the maximum sensitivity. For LISA, these two quantities take the values, $L=2.5 \times 10^9~\text{m}$ and $f_{*}=19.09~\text{mHz}$. In addition to these, there are two more functions on which the noise spectral density of LISA depends on, and these are given by,
\begin{eqnarray}
P_{\rm OMS}&=&(1.5 \times 10^{-11}~\text{m})^2 \Big[1+\Big(\dfrac{2~\text{mHz}}{f}\Big)^4\Big]~\text{Hz}^{-1}~,
\\
P_{\rm acc}&=&(1.5\times 10^{-11}~\text{m}s^{-2})^2 \Big[1+\Big(\dfrac{0.4~\text{mHz}}{f}\Big)^2\Big]\Big[1+\Big(\dfrac{f}{8~\text{mHz}}\Big)^4\Big]~.
\end{eqnarray}
Given the energy carried out by the GW, whose expression is given by \ref{hrms} and the noise spectral density $S_{\rm h}(f)$, we are fully equipped to compute the signal to noise (SNR) ratio. This ratio is essentially related with the detectablity of a particular GW event. For example, an event is likely to be detectable if its SNR is more than the threshold value of SNR of the corresponding GW detector, which is detecting the event. For a signal with duration of $\Delta t$ which corresponds to a bandwidth of $\Delta f$, the rms value of the SNR, $\rho_{\rm rms}$, is given as \cite{hawking1987three}
\begin{equation}
    \rho_{\rm rms}=\left\langle\dfrac{S}{N}\right \rangle_{\rm rms}=\dfrac{h_{\rm rms}}{\sqrt{S_h(f_{\rm isco})\Delta f}},
\end{equation}
where $h_{\rm rms}$ is the rms value of a signal, and the noise spectral density $S_{\rm h}(f)$ is computed at the ISCO. The expression for $h_{\rm rms}$ is given in \ref{hrms}, with $\Omega_{\rm isco}$ and $\dot{\mathcal{E}}$ are supplied from \ref{eq:Omegaisco} and \ref{eq:DEDT} respectively. However, the above expression excludes the average contributions from  antenna functions of the detectors. By considering them, the above expression would transform into \cite{Robson:2018ifk}:
\begin{equation}\label{eq:rhorms}
\rho^2_{\rm rms}=\dfrac{3}{10}\dfrac{h^2_{\rm rms}}{S_{\rm h}(f_{\rm isco})}\Delta t~,
\end{equation}
where the factor of $(3/10)$ comes from averaging over the antenna pattern and $\Delta t$ is the timescale associated with the transition from the inspiral to the plunging phase, and is given by the inverse of the bandwidth, i.e., $\Delta t=1/\Delta f$. We have tabulated estimates for $\Delta t$ for various choices of the rotation parameter $a$ and tidal charge $q$ in \ref{Tab_01} and \ref{Tab_02}, respectively. It should be emphasized that, unlike LIGO, for LISA, the antenna functions have explicit frequency dependence, and the factor of $(3/10)$ gives the leading contribution to the averaged antenna function. The result of the above analysis, using the energy loss formula derived in the previous section for the braneworld black holes, have been presented in \ref{fig:figure_02}. 
The left figure in \ref{fig:figure_02} demonstrates the variation of the SNR with the tidal charge parameter, for both positive and negative values of the tidal charge and also for different choices of the rotation parameter. On the other hand, the right-hand figure in \ref{fig:figure_02} depicts the variation of the SNR with the rotation parameter for different choices of the tidal charge, both positive and negative. As evident from these plots, the value of SNR largely depends on the tidal charge $q$. In particular, the nonzero and negative values of tidal charge increase the SNR value. This also implies that the energy released by GW in the transition regime is larger for braneworld BH, than for a Reissner-Nordstr\"{o}m BH and is a direct observational probe of the existence of higher spatial dimensions. This suggests that excess power in the GWs emitted during the transition regime of EMRI, identified with the result that GW frequency should be twice the angular velocity at the ISCO, signals the possible existence of higher spatial dimensions. Albeit, we use a simple setup that should be replaced with a more accurate formalism to obtain the SNR appropriately. We emphasize that our study is merely indicating the order of magnitude of SNR during the inspiral to plunge phase and how the charge is modifying these results. For practical detection purposes, these results should be modified by incorporating an accurate waveform model, introducing eccentricity and inclination in the motion, etc. Even then, it is expected that the structural dependence of the SNR on the tidal charge $q$ would be similar to that of \ref{fig:figure_02}, and the transition regime continues to be relevant to study braneworld BHs. Hopefully, as and when LISA becomes operational, it will open a new window for the search of extra spatial dimensions, or, in general for scenarios beyond general relativity.  
\begin{table}[htp]
\setlength{\tabcolsep}{10pt}
\caption[The.]{We have tabulated the timescale associated with the plunging phase for zero rotation and for a set of values for the tidal charge. These plots are produced by assuming that the inspiral starts at a distance close to the ISCO, in particular, $R_{\rm initial}=0.01 M$. We then numerically evolve \ref{eq:R_evolve} and note down the time when the secondary object crosses the ISCO. The difference is given by $\Delta t$, which we express in seconds, by scaling the mass with appropriate scaling factors, i.e., $(G M/c^3)$. Note that the negative values of the tidal charge significantly increase the timescale which would also increase their chance of getting detected, as evident from \ref{fig:figure_02}. \\}
\label{Tab_01}
\begin{center}
\begin{tabular}{ccccc}
 \hline\hline
 \multirow{5}{*}{Rotation parameter ($a/M$)}& \multicolumn{3}{c}{} \\
 \\
 & Tidal charge ($q$) & Frequency (f) & $\Delta t$ \\
 & & (in Hz) & (in sec) \\
 \\
   \hline
      &  -$0.40$ & $0.004$ & $8733$\\
 &  -$0.20$ & $0.004$ & $4966$  \\
0.0   & $0.00$ & 0.004  & 3703
   \\
   & $0.20$ & $0.005$ & $2979$
\\ 
& $0.40$ & $0.005$ & $2530$ 
\\
\hline\hline
\end{tabular}
\end{center}
\end{table}

\begin{table}[htp]
\setlength{\tabcolsep}{10pt}
\caption[The.]{We have tabulated the timescale associated with the plunging phase for a set of rotation parameters involving both co-rotating and counter-rotating cases, for a negative value of the tidal charge parameter. In these plots, we assume that the inspiral starts at a distance close to the ISCO, in particular, at $R_{\rm initial}=0.01 M$, where $R=r-r_{\rm isco}$. Evolving \ref{eq:R_evolve} numerically we determine $\Delta t$, which we express in seconds using appropriate scaling. \\}
\label{Tab_02}
\begin{center}
\begin{tabular}{ccccc}
 \hline\hline
 \multirow{5}{*}{Rotation parameter ($a/M$)}& \multicolumn{3}{c}{} \\
 \\
 & Tidal charge ($q$) & Frequency (f) & $\Delta t$ \\
 & & (in Hz) & (in sec) \\
 \\
   \hline
    -0.4  &   & $0.003$ & $5484$\\
-0.2 &   & $0.004$ & $4830$  \\
0.0 & -0.1  & $0.004$ & $4233$  \\
0.2 &   & $0.005$ & $3699$  \\
0.4 & & 0.006 & 3235 \\
\hline\hline
\end{tabular}
\end{center}
\end{table}

\section{Discussion and concluding remarks}
In the present article, we have studied the transition from the inspiral to the plunge for a braneworld EMRI, consisting of a secondary object and a primary object, whose geometry is that of a braneworld BH. By assuming that the central supermassive object is endowed with corrections from higher dimensions, we model the final stage of the binary evolution, as the secondary object inspiralling around the primary makes its transition to the final plunge phase. The aim of this article is to develop a quantitative understanding of how the existence of the tidal charge in the braneworld scenario influence the transition phase and hence the energy emitted by the GWs in this phase. Since the transition phase happens close to the ISCO, where the gravitational effects of the central primary braneworld BH is strong enough, we expect the GWs emitted during this phase to encode significant information about the spacetime geometry, and as we have demonstrated this is indeed the case. Intriguingly, the modifications due to the braneworld scenario considered here, is akin to the Kerr-Newman spacetime for a certain parameter range of the tidal charge and also some other metric solutions in Horndeski theories \cite{Mukherjee:2017fqz}. Therefore, the results presented in this work are expected to be valid for a larger spectrum of gravity theories beyond the general relativity. 

In the first part of the work, we introduce the orbital dynamics in the braneworld gravity, with emphasis on the circular geodesics and their properties. We then evaluate the fluxes for the energy, the angular momentum and the Carter constant. These fluxes are evaluated by ignoring $\mathcal{O}(a^2,q^2,aq)$ terms of the central BH, which makes the analysis valid for slowly rotating and weakly tidally charged BHs. Expressions for fluxes are obtained in the case of off-equatorial circular/spherical orbits. Interestingly, our results demonstrate that the contribution in the above fluxes from the tidal charge appears at a lower PN order than the contribution from the rotation of the BH. This opens up the tantalizing possibility, that even if the tidal charge parameter is an order of magnitude smaller than the rotation parameter of the black hole, it actually contributes equally to the energy loss through GWs and hence is detectable.

In the second part of the work, we have used the expression for the energy flux in order to find out the energy carried out in the form of GW during the transition regime. The transition regime is important as it bridges between the inspiral and the plunge phase, and since this phase happens close to the ISCO, it is expected that signatures from the strong gravity regime, e.g., that of the tidal charge will be embedded into the signal. The effect will only be appreciable for EMRI, since then the time scale the secondary object spends during the transition regime will be significant. Following this, we indeed found that the transition regime is significantly affected due to the tidal charge, and most importantly, on its sign.  For example, with $q>0$, the transition to the plunge phase happens much faster and the effect of the tidal charge on the GW waveform will be less pronounced. On the contrary, for $q<0$, the transition to the plunge phase happens at a much slower rate and it takes a longer time to reach even to the ISCO. Therefore, any effect of the negative tidal charge will be enhanced by the GW waveform as it emits a larger amount of energy through GWs. These features are reflected in \ref{fig:figure_01}. The signal to noise ratio $\rho_{\rm rms}$, also tells a similar story. Existence of negative tidal charge and motion along the direction of rotation of the black hole enhances the signal to noise ratio. This suggests that, if the braneworld BHs are indeed realized in nature and have negative tidal charge, they will contribute more to the energy flux in the form of GWs, than their general relativistic counterpart. Thus, EMRI systems and in particular the transition phase is one of the promising avenue to look for any signatures of the deviations from general relativity.  

Finally, we should conclude by discussing the possible extensions of this work. First of all, the fluxes derived in \ref{eq:DEDT} to \ref{eq:DQDT} are based on slow rotation and small tidal charge approximations. It will be interesting to generalize the setup presented here, to more accurate flux measurement techniques, using e.g., the \textit{Teukolsky formalism} \cite{Teukolsky:1973ha,Hughes:1999bq}. Moreover, the analysis presented for the transition regime is dependent on orbits which are equatorial and circular. Possible generalization to non-equatorial, eccentric orbit may provide better understanding of the transition regime and the energy loss by GWs. This is important, since typical EMRIs will have both eccentricity and inclination when entering the LISA band and should be modelled with generic orbits \cite{Drasco:2005kz}. We hope to come back to these questions in a future work.

\section*{Acknowledgements}

Research of S.C. is funded by the INSPIRE Faculty fellowship from DST, Government of India (Reg. No. DST/INSPIRE/04/2018/000893) and by the Start-Up Research Grant from SERB, DST, Government of India (Reg. No. SRG/2020/000409). S.M. is supported by the fellowship Lumina Quaeruntur No. LQ100032102 of the Czech Academy of Sciences. Finally, the authors are thankful to  Georgios Lukes-Gerakopoulos for reading the manuscript and giving constructive suggestions and comments. 

\appendix
\labelformat{section}{Appendix #1} 
\section{Derivation of the ISCO for rotating braneworld spacetime}\label{AppA}

In this appendix, we will provide the detailed derivation of the location of the ISCO for the rotating braneworld spacetime. For this purpose, we recall the expressions of the conserved energy and the conserved angular momentum of a particle in a circular orbit on the equatorial plane of the rotating braneworld black hole spacetime, these are given by, \ref{circ_E} and \ref{circ_L}, respectively. From these, we can work out the following qunatity,
\begin{widetext}
\begin{align}
aE_{\rm (circ)}-L_{z~\textrm{(circ)}}&=a\frac{1-\frac{2M}{r}+\frac{qM^{2}}{r^{2}}\pm \frac{a}{r}\sqrt{\frac{M}{r}-\frac{qM^{2}}{r^{2}}}}{\sqrt{1-\frac{3M}{r}+\frac{2qM^{2}}{r^{2}}\pm 2\frac{a}{r}\sqrt{\left(\frac{M}{r}-\frac{qM^{2}}{r^{2}}\right)}}}\mp \frac{\left(1+\frac{a^{2}}{r^{2}}\right)\sqrt{Mr-qM^{2}}\mp a\left(\frac{2M}{r}-\frac{qM^{2}}{r^{2}}\right)}{\sqrt{1-\frac{3M}{r}+\frac{2qM^{2}}{r^{2}}\pm 2\frac{a}{r}\sqrt{\left(\frac{M}{r}-\frac{qM^{2}}{r^{2}}\right)}}}
\nonumber
\\
&=\frac{a-a\left(\frac{2M}{r}-\frac{qM^{2}}{r^{2}}\right)\pm \frac{a^{2}}{r}\sqrt{\frac{M}{r}-\frac{qM^{2}}{r^{2}}}\mp\left(1+\frac{a^{2}}{r^{2}}\right)\sqrt{Mr-qM^{2}}+ a\left(\frac{2M}{r}-\frac{qM^{2}}{r^{2}}\right)}{\sqrt{1-\frac{3M}{r}+\frac{2qM^{2}}{r^{2}}\pm 2\frac{a}{r}\sqrt{\left(\frac{M}{r}-\frac{qM^{2}}{r^{2}}\right)}}}
\nonumber
\\
&=\frac{a\mp\sqrt{Mr-qM^{2}}}{\sqrt{1-\frac{3M}{r}+\frac{2qM^{2}}{r^{2}}\pm 2\frac{a}{r}\sqrt{\left(\frac{M}{r}-\frac{qM^{2}}{r^{2}}\right)}}}
\end{align}
\end{widetext}
This expression is helpful, since it appears in the relation determine the innermost stable circular orbit (ISCO), as evident from \ref{isco_rel} in the main text. Further simplification yields,
\begin{widetext}
\begin{align}
0&=-2+2\frac{\left(1-\frac{2M}{r}+\frac{qM^{2}}{r^{2}}\pm \frac{a}{r}\sqrt{\frac{M}{r}-\frac{qM^{2}}{r^{2}}}\right)^{2}}{1-\frac{3M}{r}+\frac{2qM^{2}}{r^{2}}\pm 2\frac{a}{r}\sqrt{\left(\frac{M}{r}-\frac{qM^{2}}{r^{2}}\right)}}+\left(\frac{4M}{r^{3}}-\frac{6qM^{2}}{r^{4}}\right)\frac{\left(a\mp\sqrt{Mr-qM^{2}}\right)^{2}}{1-\frac{3M}{r}+\frac{2qM^{2}}{r^{2}}\pm 2\frac{a}{r}\sqrt{\left(\frac{M}{r}-\frac{qM^{2}}{r^{2}}\right)}}
\nonumber
\\
&=\frac{-2+\frac{6M}{r}-\frac{4qM^{2}}{r^{2}}\mp 4\frac{a}{r}\sqrt{\left(\frac{M}{r}-\frac{qM^{2}}{r^{2}}\right)}+2\left(1-\frac{2M}{r}+\frac{qM^{2}}{r^{2}}\right)^{2}\pm 4\left(1-\frac{2M}{r}+\frac{qM^{2}}{r^{2}}\right)\frac{a}{r}\sqrt{\frac{M}{r}-\frac{qM^{2}}{r^{2}}}}{1-\frac{3M}{r}+\frac{2qM^{2}}{r^{2}}\pm 2\frac{a}{r}\sqrt{\left(\frac{M}{r}-\frac{qM^{2}}{r^{2}}\right)}}
\nonumber
\\
&\hskip 2 cm +\frac{2\frac{a^{2}}{r^{2}}\left(\frac{M}{r}-\frac{qM^{2}}{r^{2}}\right)+\left(\frac{4M}{r^{3}}-\frac{6qM^{2}}{r^{4}}\right)\left(a^{2}+Mr-qM^{2}\mp 2a\sqrt{Mr-qM^{2}} \right)}{1-\frac{3M}{r}+\frac{2qM^{2}}{r^{2}}\pm 2\frac{a}{r}\sqrt{\left(\frac{M}{r}-\frac{qM^{2}}{r^{2}}\right)}}
\nonumber
\\
&=\frac{-2+\frac{6M}{r}-\frac{4qM^{2}}{r^{2}}+2\left(1-\frac{4M}{r}+\frac{4M^{2}}{r^{2}}+\frac{q^{2}M^{4}}{r^{4}}+\frac{2qM^{2}}{r^{2}}-\frac{4qM^{3}}{r^{3}}\right)\pm 4\left(-\frac{2M}{r}+\frac{qM^{2}}{r^{2}}\right)\frac{a}{r}\sqrt{\frac{M}{r}-\frac{qM^{2}}{r^{2}}}}{1-\frac{3M}{r}+\frac{2qM^{2}}{r^{2}}\pm 2\frac{a}{r}\sqrt{\left(\frac{M}{r}-\frac{qM^{2}}{r^{2}}\right)}}
\nonumber
\\
&\hskip 2 cm +\frac{\frac{6Ma^{2}}{r^{3}}-\frac{8qM^{2}a^{2}}{r^{4}}+\frac{4M^{2}}{r^{2}}-\frac{10qM^{3}}{r^{3}}+\frac{6q^{2}M^{4}}{r^{4}}\mp 2a\sqrt{Mr-qM^{2}} 
\left(\frac{4M}{r^{3}}-\frac{6qM^{2}}{r^{4}}\right)}{1-\frac{3M}{r}+\frac{2qM^{2}}{r^{2}}\pm 2\frac{a}{r}\sqrt{\left(\frac{M}{r}-\frac{qM^{2}}{r^{2}}\right)}}
\nonumber
\\
&=\frac{-\frac{2M}{r}+\frac{8M^{2}}{r^{2}}+\frac{2q^{2}M^{4}}{r^{4}}-\frac{8qM^{3}}{r^{3}}\pm 4\left(-\frac{2M}{r}+\frac{qM^{2}}{r^{2}}\right)\frac{a}{r}\sqrt{\frac{M}{r}-\frac{qM^{2}}{r^{2}}}}{1-\frac{3M}{r}+\frac{2qM^{2}}{r^{2}}\pm 2\frac{a}{r}\sqrt{\left(\frac{M}{r}-\frac{qM^{2}}{r^{2}}\right)}}
\nonumber
\\
&\hskip 2 cm +\frac{\frac{6Ma^{2}}{r^{3}}-\frac{8qM^{2}a^{2}}{r^{4}}+\frac{4M^{2}}{r^{2}}-\frac{10qM^{3}}{r^{3}}+\frac{6q^{2}M^{4}}{r^{4}}\mp 2\frac{a}{r}\sqrt{\frac{M}{r}-\frac{qM^{2}}{r^{2}}} \left(\frac{4M}{r}-\frac{6qM^{2}}{r^{2}}\right)}{1-\frac{3M}{r}+\frac{2qM^{2}}{r^{2}}\pm 2\frac{a}{r}\sqrt{\left(\frac{M}{r}-\frac{qM^{2}}{r^{2}}\right)}}
\nonumber
\\
&=\frac{-\frac{2M}{r}+\frac{12M^{2}}{r^{2}}+\frac{8q^{2}M^{4}}{r^{4}}-\frac{18qM^{3}}{r^{3}}+\frac{6Ma^{2}}{r^{3}}-\frac{8qM^{2}a^{2}}{r^{4}}\pm 2\left(-\frac{8M}{r}+\frac{8qM^{2}}{r^{2}}\right)\frac{a}{r}\sqrt{\frac{M}{r}-\frac{qM^{2}}{r^{2}}}}{1-\frac{3M}{r}+\frac{2qM^{2}}{r^{2}}\pm 2\frac{a}{r}\sqrt{\left(\frac{M}{r}-\frac{qM^{2}}{r^{2}}\right)}}
\nonumber
\\
&=\frac{-\frac{2M}{r^{4}}\left(r^{3}-6Mr^{2}-4q^{2}M^{3}+9qM^{2}r-3a^{2}r+4qa^{2}M^{3}\pm 8\left(r-qM\right)a\sqrt{Mr-qM^{2}}\right)}{1-\frac{3M}{r}+\frac{2qM^{2}}{r^{2}}\pm 2\frac{a}{r}\sqrt{\left(\frac{M}{r}-\frac{qM^{2}}{r^{2}}\right)}}
\end{align}
\end{widetext}
From this result, the algebraic expression in the main text, yielding the location of the innermost stable circular orbit follows. In addition to the above relation, we also note down the expression of the angular velocity at the ISCO, denoted by $\Omega_{\rm isco}$, to be given by,
\begin{equation}
    \Omega_{\rm isco}=M^{1/2}\dfrac{\sqrt{r_{\rm isco}-q M}}{r^2_{\rm isco}+a M^{1/2}\sqrt{r_{\rm isco}-qM}}~.
    \label{eq:OMEGA_ISCO}
\end{equation}
which will be used in the main text. 

\section{Derivation of the Lagrangian for a particle in rotating braneworld spacetime}\label{AppB}

In this appendix, we will derive the effective Lagrangian for a particle moving in the rotating braneworld black hole spacetime. For this purpose, we will employ the Hamilton-Jacobi formalism, for which the action can be written as,
\begin{align}
\mathcal{A}=-Et+L_{z}\phi+R(r)+\Theta(\theta)
\end{align}
From the equation, $g^{\mu \nu}\partial_{\mu}\mathcal{A}\partial_{\nu}\mathcal{A}=-m^{2}$, we obtain the following two equations, 
\begin{align}
\left(\frac{d\Theta}{d\theta}\right)^{2}&+\left(aE\sin\theta-\frac{L_{z}}{\sin \theta}\right)^{2}+a^{2}m^{2}\cos^{2}\theta=K
\label{action_radial}
\\
\Delta \left(\frac{dR}{dr}\right)^{2}&-\frac{1}{\Delta}\left[\left(r^{2}+a^{2}\right)E-aL_{z}\right]^{2}+m^{2}r^{2}=-K
\label{action_angle}
\end{align}
Here, $K$ is referred to as the Carter's constant. Again, the above quantities $(dR/dr)$ and $(d\Theta/d\theta)$ can be related to the radial and angular momentum, such that,
\begin{align}
\frac{dR}{dr}&=m\frac{\rho^{2}}{\Delta}\left(\frac{dr}{d\tau}\right)
\\
\frac{d\Theta}{d\theta}&=m\rho^{2}\left(\frac{d\theta}{d\tau}\right)
\end{align}
Thus, \ref{action_radial} and \ref{action_angle} can be expressed as, 
\begin{align}
m^{2}\left(\frac{dr}{d\tau}\right)^{2}&=\frac{1}{\rho^{4}}\left[\left(r^{2}+a^{2}\right)E-aL_{z}\right]^{2}-\frac{\Delta}{\rho^{4}}\left(K+m^{2}r^{2}\right)
\\
m^{2}\left(\frac{d\theta}{d\tau}\right)^{2}&=\frac{1}{\rho^{4}}\left(K-a^{2}m^{2}\cos^{2}\theta\right)
-\frac{1}{\rho^{4}}\left(aE\sin\theta-\frac{L_{z}}{\sin \theta}\right)^{2}
\end{align}
Along with these relations, we also have two additional ones, arising from $(\partial \mathcal{A}/\partial t)=-E$ and $(\partial \mathcal{A}/\partial \phi)=L_{z}$. Again, the conserved energy and angular momentum can be written in terms of $(dt/d\tau)$ and $(d\phi/d\tau)$, such that, 
\begin{align}
m\left(\frac{dt}{d\tau}\right)&=-\left[\left(r^{2}+a^{2}\right)-\Delta\right]\frac{aL_{z}}{\rho^{2}\Delta}+\left[\left(r^{2}+a^{2}\right)^{2}-\Delta a^{2}\sin^{2}\theta\right]\frac{E}{\rho^{2}\Delta}
\\
m\left(\frac{d\phi}{d\tau}\right)&=\left(1-\frac{r^{2}+a^{2}-\Delta}{\rho^{2}}\right)\frac{L_{z}}{\Delta \sin^{2}\theta}+\left[\left(r^{2}+a^{2}\right)-\Delta\right]\frac{aE}{\rho^{2}\Delta}
\end{align}
In the limit of small rotation, ignoring all terms of $\mathcal{O}(a^{2})$, we obtain the following expressions for the velocity components of a geodesic, with mass $m$, moving in the rotating braneworld black hole spacetime, 
\begin{align}
m\left(\frac{dt}{d\tau}\right)&=-\left(r^{2}-\Delta\right)\frac{aL_{z}}{r^{2}\Delta}+\frac{r^{2}E}{\Delta}
\label{dtdtau}
\\
m\left(\frac{d\phi}{d\tau}\right)&=\frac{L_{z}}{r^{2}\sin^{2}\theta}+\frac{aE}{r^{2}\Delta}\left(r^{2}-\Delta\right)
\label{dphidtau}
\\
m^{2}\left(\frac{dr}{d\tau}\right)^{2}&=\frac{1}{r^{4}}\left(r^{2}E-aL_{z}\right)^{2}-\frac{\Delta}{r^{4}}\left(K+m^{2}r^{2}\right)
\\
m^{2}\left(\frac{d\theta}{d\tau}\right)^{2}&=\frac{K}{r^{4}}-\frac{1}{r^{4}}\left(aE\sin\theta-\frac{L_{z}}{\sin \theta}\right)^{2}
\label{dthetadtau}
\end{align}
Here, $\Delta=r^{2}-2Mr+M^{2}q$, since all terms of $\mathcal{O}(a^{2})$ are being neglected. Given the above components of four-velocity for geodesic motion in the spacetime, we can construct the following combination, 
\begin{align}
m^{2}\Bigg[\left(\frac{dr}{d\tau}\right)^{2}&+r^{2}\left(\frac{d\theta}{d\tau}\right)^{2}+r^{2}\sin^{2}\theta\left(\frac{d\phi}{d\tau}\right)^{2}\Bigg]
\nonumber
\\
&=E^{2}-\frac{2aEL_{z}}{r^{2}}-\frac{\Delta}{r^{4}}\left(K+m^{2}r^{2}\right)+\frac{K}{r^{2}}-\frac{1}{r^{2}}\left(aE\sin\theta-\frac{L_{z}}{\sin \theta}\right)^{2}
\nonumber
\\
&+r^{2}\sin^{2}\theta\left[\frac{L_{z}}{r^{2}\sin^{2}\theta}+\frac{aE}{r^{2}\Delta}\left(r^{2}-\Delta\right)\right]^{2}
\nonumber
\\
&=E^{2}-\frac{\Delta K}{r^{4}}-\frac{m^{2}\Delta}{r^{2}}+\frac{K}{r^{2}}+2aEL_{z}\frac{\left(2Mr-M^{2}q\right)}{r^{2}\left(r^{2}-2Mr+M^{2}q\right)}
\nonumber
\\
&=E^{2}+\frac{2MK}{r^{3}}-\frac{KM^{2}q}{r^{4}}-m^{2}+\frac{2Mm^{2}}{r}-\frac{m^{2}M^{2}q}{r^{2}}
\end{align}
where, we have neglected terms of $\mathcal{O}(aM^{2}/r^{3})$ and $\mathcal{O}(aqM^{3}/r^{4})$. We can express all the derivatives with respect to the proper time $\tau$ to derivatives with respect to the Boyer-Lindquist coordinate time $t$, using $(dt/d\tau)$ from \ref{dtdtau}. Again ignoring terms of $\mathcal{O}(r^{3})$ and higher, we obtain, 
\begin{align} \label{eq:Hamiltonian}
m^{2}\Bigg[\left(\frac{dr}{dt}\right)^{2}&+r^{2}\left(\frac{d\theta}{dt}\right)^{2}+r^{2}\sin^{2}\theta\left(\frac{d\phi}{dt}\right)^{2}\Bigg]
\nonumber
\\
&=\left(E^{2}-m^{2}+\frac{2Mm^{2}}{r}-\frac{m^{2}M^{2}q}{r^{2}}\right)\left(\frac{m\Delta}{Er^{2}}\right)^{2}
\end{align}
Finally, writing, $E=m+\mathcal{E}$, i.e., separating out the rest mass contribution, we obtain, $E^{2}-m^{2}=2m\mathcal{E}+\mathcal{O}(\mathcal{E}^{2})$ and $(m/E)=1-(\mathcal{E}/m)+\mathcal{O}(\mathcal{E}^{2})$. Using these results and ignoring all the terms $\mathcal{O}(\mathcal{E}^{2})$, as well as terms $\mathcal{O}(\mathcal{E}M/r)$, $\mathcal{O}(\mathcal{E}q/r^{2})$, we obtain the following expression for the energy $\mathcal{E}$ of the object, 
\begin{align}
\mathcal{E}=\frac{m}{2}\Bigg[\left(\frac{dr}{dt}\right)^{2}&+r^{2}\left(\frac{d\theta}{dt}\right)^{2}+r^{2}\sin^{2}\theta\left(\frac{d\phi}{dt}\right)^{2}\Bigg]-\frac{mM}{r}+\frac{mM^{2}q}{2r^{2}}
\label{energy}
\end{align}
This completes one part of the story. For the other, we start with $(d\phi/d\tau)$ and $(d\theta/d\tau)$ from \ref{dphidtau} and \ref{dthetadtau}, respectively. Squaring these quantities, then multiplying appropriate factors and finally adding these up, we arrive at the following relation,
\begin{align}
m^{2}r^{4}\Bigg[\left(\frac{d\theta}{d\tau}\right)^{2}&+\sin^{2}\theta\left(\frac{d\phi}{d\tau}\right)^{2}\Bigg]=K-\left(aE\sin\theta-\frac{L_{z}}{\sin \theta}\right)^{2}+\frac{1}{\sin^{2}\theta}\left[L_{z}+\frac{aE\sin^{2}\theta}{\Delta}\left(2Mr-M^{2}q\right)\right]^{2}
\nonumber
\\
&\simeq K+2aEL_{z}\left(1+\frac{2Mr-M^{2}q}{\Delta}\right)
\nonumber
\\
&\simeq K+2aEL_{z}\left(1+\frac{2M}{r}-\frac{M^{2}q}{r^{2}}\right)
\end{align}
Since, the right hand side is already linear in the rotation parameter $a$ along with the Carter constant $K$, changing proper time $\tau$ to coordinate time $t$ can be done without introducing any additional factors. Also the term involving $(aq/r^{2})$ can be neglected, since both the rotation and the tidal charge is taken to be small. Therefore, we arrive at, 
\begin{align}
m^{2}r^{4}\Bigg[\left(\frac{d\theta}{dt}\right)^{2}&+\sin^{2}\theta\left(\frac{d\phi}{dt}\right)^{2}\Bigg]=\underbrace{K+2aEL_{z}}_{Q+L_{z}^{2}}+\frac{4aMEL_{z}}{r}
\label{intermediate}
\end{align}
Here, we have introduced a more conventional Carter constant $Q$, through the relation, $Q+L_{z}^{2}=K+2aEL_{z}$. Finally, from \ref{dphidtau}, we can express the conserved angular momentum as, 
\begin{align}
L_{z}=mr^{2}\sin^{2}\theta\left(\frac{d\phi}{d\tau}\right)-\frac{2MaE\sin^{2}\theta}{r}
\label{Lz}
\end{align}
Substitution of the conserved angular momentum $L_{z}$ from \ref{Lz} in \ref{intermediate}, and ignoring any terms $\mathcal{O}(a^{2})$, we obtain, 
\begin{align}
m^{2}r^{4}\Bigg[\left(\frac{d\theta}{dt}\right)^{2}&+\sin^{2}\theta\left(\frac{d\phi}{dt}\right)^{2}\Bigg]=Q+L_{z}^{2}
+4am^{2}Mr\sin^{2}\theta\left(\frac{d\phi}{dt}\right)
\label{thetaphi}
\end{align}
This can also be written in the following form, 
\begin{align}
Q+L_{z}^{2}=m^{2}r^{4}\Bigg[\left(\frac{d\theta}{dt}\right)^{2}&+\sin^{2}\theta\left(\frac{d\phi}{dt}\right)^{2}\Bigg]
-4am^{2}Mr\sin^{2}\theta\left(\frac{d\phi}{dt}\right)
\end{align}
Here, we have changed $E$ to $m$ and have changed the proper time $\tau$ to coordinate time $t$, since the conserved angular momentum $L_{z}$ is already multiplied by the rotation parameter $a$. Note that this expression has been used in the main text. Given all these results, we can find out the effective Lagrangian $L_{\rm eff}$. First of all, we consider \ref{Lz} and note that $L_{z}=[\partial L_{\rm eff}/\partial\dot{\phi}]$, where `dot' denotes derivative with respect to the coordinate time $t$. This suggests that the effective Lagrangian should have the form, 
\begin{align}
L_{\rm eff}=\frac{m}{2}r^{2}\sin^{2}\theta\left(\frac{d\phi}{dt}\right)^{2}-\frac{2maM\sin^{2}\theta}{r}\left(\frac{d\phi}{dt}\right)+f(r,\theta,\dot{r},\dot{\theta})
\label{appeffL}
\end{align}
We now have the following relation between the effective Lagrangian $L_{\rm eff}$ and the Hamiltonian (which is the same as the energy $\mathcal{E}$), 
\begin{align}
\mathcal{E}&=\left(\frac{\partial L_{\rm eff}}{\partial\dot{r}}\right)\dot{r}+\left(\frac{\partial L_{\rm eff}}{\partial\dot{\theta}}\right)\dot{\theta}+\left(\frac{\partial L_{\rm eff}}{\partial\dot{\phi}}\right)\dot{\phi}-L_{\rm eff}
\nonumber
\\
&=\left(\frac{\partial f}{\partial\dot{r}}\right)\dot{r}+\left(\frac{\partial f}{\partial\dot{\theta}}\right)\dot{\theta}
+\left[mr^{2}\sin^{2}\theta\dot{\phi}-\frac{2maM\sin^{2}\theta}{r}\right]\dot{\phi}
-\frac{m}{2}r^{2}\sin^{2}\theta\dot{\phi}^{2}
\nonumber
\\
&\hskip 2 cm +\frac{2maM\sin^{2}\theta}{r}\dot{\phi}-f(r,\theta,\dot{r},\dot{\theta})
\nonumber
\\
&=\left(\frac{\partial f}{\partial\dot{r}}\right)\dot{r}+\left(\frac{\partial f}{\partial\dot{\theta}}\right)\dot{\theta}
+\frac{m}{2}r^{2}\sin^{2}\theta\dot{\phi}^{2}-f(r,\theta,\dot{r},\dot{\theta})
\end{align}
This demands, given \ref{energy}, the following form for the function $f(r,\theta,\dot{r},\dot{\theta})$ 
\begin{align}
f(r,\theta,\dot{r},\dot{\theta})=\frac{m}{2}\dot{r}^{2}+\frac{m}{2}r^{2}\dot{\theta}^{2}+\frac{mM}{r}-\frac{mM^{2}q}{2r^{2}}
\end{align}
and hence the effective Lagrangian can be determined by the substitution of the function $f$ into \ref{appeffL}, which will lead to \ref{effL} in the main text. 

\section{Solution of the radial differential equation}\label{AppC}

In this appendix, we will provide a general solution to the differential equation for the radial motion, presented in \ref{eq:rddot2}. For this purpose, we note that any second order differential equation of the following form can be solved exactly:
\begin{equation}
\ddot{r}+\dfrac{A}{r^2}-\dfrac{B}{r^3}-\dfrac{C}{r^4}=0~,
\end{equation}
with the solution being given by,
\begin{equation}
r=\Big(\dfrac{B}{A}\Big)\dfrac{1}{1+e \cos\psi}\Big[1+\dfrac{AC}{B^2}\left\{1+\dfrac{1}{3}e \cos\psi\right\}\Big]~.
\end{equation}
In the above, the term $e$ is the eccentricity of the orbit, and we define $\psi$ as 
\begin{equation}
\dfrac{dt}{d\psi}=\Big(\dfrac{B\sqrt{B}}{A^2}\Big)\dfrac{1}{(1+e\cos\psi)^2}\Big[1+\dfrac{AC}{B^2}\Big]   
\end{equation}
Choosing, $A=M$, $B=qM^{2}+\{(Q+L_{z}^{2})/m^{2}\}$ and $C=(6aML_{z}/m)$, we arrive at the solution of the radial coordinate in the main text. 

\section{Alternative derivation of the circular orbit}\label{Appd}

In this appendix, we will present an alternative derivation of the radius of the circular orbit, starting from \ref{eq:rddot2}. First, consider the case of Kerr spacetime, for which setting $q=0$ and $\ddot{r}=0$ in \ref{eq:rddot2}, we obtain, 
\begin{align}
\frac{L_{z}^{2}}{m^{2}r^{3}}-\dfrac{M}{r^2}+\dfrac{6aML_{z}}{mr^4}=0~.
\end{align}
This can be translated to the following algebraic equation, 
\begin{align}
\dfrac{6aML_{z}}{m}+\frac{L_{z}^{2}}{m^{2}}r-Mr^{2}=0~,
\end{align}
with the following solution, 
\begin{align}
r_{\rm Kerr}&=\left(-\frac{1}{2M}\right)\left[-\frac{L_{z}^{2}}{m^{2}}-\sqrt{\frac{L_{z}^{4}}{m^{4}}+\dfrac{24aM^{2}L_{z}}{m}}\right]
\nonumber
\\
&=\left(\frac{1}{2M}\right)\left[\frac{L_{z}^{2}}{m^{2}}+\frac{L_{z}^{2}}{m^{2}}\left(1+\dfrac{12am^{3}M^{2}}{L_{z}^{3}}\right)\right]
\nonumber
\\
&=\frac{L_{z}^{2}}{Mm^{2}}+\frac{6amM}{L_{z}}
\end{align}
As evident, this coincides with the $q=0$ limit of \ref{rcirc} in the main text. Choosing the ansatz $r=r_{\rm Kerr}+g(q,a)$, as in the main text, we observe that $g$ will satisfy the following quadratic equation, 
\begin{align}
\dfrac{6aML_{z}}{m}+\frac{L_{z}^{2}+qM^{2}m^{2}}{m^{2}}\left(r_{\rm Kerr}+g\right)-M\left(r_{\rm Kerr}+g\right)^{2}=0~.
\end{align}
with the following solution, 
\begin{align}
g&=\left(-\frac{1}{2M}\right)\left[-\frac{L_{z}^{2}+qM^{2}m^{2}}{m^{2}}
-\sqrt{\left(\frac{L_{z}^{2}+qM^{2}m^{2}}{m^{2}}\right)^{2}+\dfrac{24aM^{2}L_{z}}{m}}\right]-r_{\rm Kerr}
\nonumber
\\
&=\frac{L_{z}^{2}+qM^{2}m^{2}}{m^{2}M}+\dfrac{6aML_{z}m}{L_{z}^{2}+qM^{2}m^{2}}-r_{\rm Kerr}
\end{align}
Here the `-'ve sign has been chosen before the square root, since we want $g=0$, in the limit $q\rightarrow 0$. Further, substituting for $r_{\rm Kerr}$, we obtain, 
\begin{align}
g&=qM+\dfrac{6aML_{z}m}{L_{z}^{2}+qM^{2}m^{2}}-\frac{6amM}{L_{z}}
\nonumber
\\
&=qM+\dfrac{6aML_{z}^{2}m-6amM\left(L_{z}^{2}+qM^{2}m^{2}\right)}{L_{z}\left(L_{z}^{2}+qM^{2}m^{2}\right)}
\nonumber
\\
&=q\left[M-\dfrac{6aM^{3}m^{3}}{L_{z}\left(L_{z}^{2}+qM^{2}m^{2}\right)} \right]
\end{align}
Note that, in the limit $q=0$, we get back the result $g=0$, as expected. Hence adding the expression for $r_{\rm Kerr}$, derived above, with the $g$ given here, we will recover, 
\begin{eqnarray}
r&=&\dfrac{L^2_{\rm z}}{m^2 M}+\dfrac{6amM}{L_{\rm z}}+q \Big[M-\dfrac{6aM^3m^3}{L_z(L^2_z+qm^2M^2)}\Big] \nonumber 
\\
&=&\dfrac{L^2_{\rm z}}{m^2 M}+qM+\dfrac{6amM}{L_{\rm z}}\Big(1-\frac{qm^2M^2}{(L_{\rm z}^2+qm^2M^2)}\Big) \nonumber 
\\
&=& \dfrac{L^2_{\rm z}}{m^2 M}+q M+\dfrac{6aMm L_{\rm z}}{L^2_{z\rm }+qm^2 M^2}~,
\end{eqnarray}
This is what yields the circular orbit radius $r_{\rm circ}$ in the main text. 
\bibliographystyle{utphys1}
\bibliography{References}
\end{document}